\documentclass[
  twocolumn,
  prb,
  showpacs,
  amsmath,
  amssymb,
  superscriptaddress
]{revtex4}

\usepackage{bm}
\usepackage{bbm}
\usepackage{graphicx}
\newcommand{\sign}{\mathop{\mathrm{sign}}}
\newcommand{\tr}{\mathop{\mathrm{tr}}}
\newcommand{\smallO}[1]{\ensuremath{o\left(#1\right)}}
\begin{document}

\title{
%\hspace*{-0.6cm} 
Correlations in non-equilibrium Luttinger liquid
and singular Fredholm determinants}

\author{I.\ V.\ Protopopov}
\affiliation{
 Institut f\"ur Nanotechnologie, Karlsruhe Institute of Technology,
 76021 Karlsruhe, Germany
}
\affiliation{
 L.\ D.\ Landau Institute for Theoretical Physics RAS,
 119334 Moscow, Russia
}

\author{D.\ B.\ Gutman}
\affiliation{Department of Physics, Bar Ilan University, Ramat Gan 52900,
Israel }
\affiliation{
 Institut f\"ur Nanotechnologie, Karlsruhe Institute of Technology,
 76021 Karlsruhe, Germany
}

\author{A.\ D.\ Mirlin}

\affiliation{
 Institut f\"ur Nanotechnologie, Karlsruhe Institute of Technology,
 76021 Karlsruhe, Germany
}
\affiliation{
 Institut f\"ur Theorie der Kondensierten Materie and DFG Center for Functional
Nanostructures,
 Karlsruhe Institute of Technology, 76128 Karlsruhe, Germany
}

%\affiliation{
% DFG Center for Functional Nanostructures, Karlsruhe Institute of Technology,
% 76128 Karlsruhe, Germany
%}

\affiliation{
 Petersburg Nuclear Physics Institute,  188300 St.~Petersburg, Russia.
}

\begin{abstract}
We study interaction-induced correlations in Luttinger liquid
with multiple Fermi edges. Many-particle correlation functions are
expressed in terms of Fredholm determinants ${\rm
det}(1+\hat{A}\hat{B})$, where $A(\epsilon)$ and $B(t)$ have multiple
discontinuities in energy and time spaces. Such determinants are a
generalization of Toeplitz determinants with Fisher-Hartwig singularities. We 
propose a general asymptotic formula for this class of
determinants and provide analytical and numerical support to this
conjecture. This allows us to establish non-equilibrium power-law
singularities of many-particle correlation functions.
As an example, we calculate a two-particle  distribution function 
characterizing correlations between left- and right-moving fermions that have
left the interaction region. 
\end{abstract}


\maketitle
%Interactions  induce correlations in the many body states, and give rise
%to a spectacular effects in  condesed matter physics.
Non-equilibrium phenomena in (effectively) one-dimensional correlated systems% 
---including Kondo and related quantum impurity models 
\cite{goldhaber-gordon,rosch,de-franceschi02,leturcq05,paaske06,mitra07,carr11},
quantum Hall edge interferometry
\cite{MachZehnder,interferometr,Litvin,Bieri,MZI-theory} and energy relaxation 
\cite{altimiras10,kovrizhin11}, carbon nanotube tunneling spectroscopy
\cite{Chen09}, Fermi edge singularity \cite{abanin04}, and correlated
electrons in quantum wires
\cite{jakobs07,gutman08,GGM_long2010,GGM_short2010,Gutman10,
NgoDinh,Protopopov2012}---%
are attracting lots of experimental  and theoretical interest.
%Out-of-equilibrium physics of 
%has been explored. 
In these problems applied voltages lead to
formation of distribution functions with two or more Fermi edges,
which generates non-equilibrium scaling, criticality and decoherence. 
%
%It is usally the case that only  limited information about the state of
%interacting electrons 
%can be extracted, even in the "simple" cases.
%The non-equilibrium LL   is notable exception. 
%the effects of interactions can be studied to the  very end and 
%
A remarkable property of the Luttinger liquid (LL) model is a possibility of
exact solution even for such non-equilibrium distributions.  
In this paper we show how many-particle fermionic correlation functions
$G_n=\langle\Psi^\dagger_1 \dots \Psi^\dagger_n\Psi_{n+1}  \dots \Psi_{2n}
\rangle$
can be calculated
%analyze the results, 
and discuss the underlying physics. It turns out that $G_n$ 
are given by a certain  type of singular Fredholm determinants.
Our result for the asymptotics of such
determinants is expected to be relevant also to other non-equilibrium many-body
problems.  

In previous works \cite{GGM_long2010,GGM_short2010,Gutman10}, we have shown
that single-particle Green function (GF) in  the case of LL (and in a number of
related problems) 
%Generalizing the  celebrated bosonization approach \cite{stone, Gogolin,
%giamarchi} we have shown that %various fermionic correlation functions these
%models 
can be expressed through  Fredholm determinants of a ``counting'' operator 
\begin{equation}
 \Delta\left[A, B\right]=\det (1+\hat{A}\hat{B}),
\label{DetDef}
\end{equation}
where operators $A(t)=e^{-i\delta(t)}{-}1$ and $B(\epsilon)=n(\epsilon)$ 
are diagonal in the time and energy representations, respectively, and
$[\hat{\epsilon},\hat{t}] = i$.
The  one-particle distribution function $n(\epsilon)$ characterizes  the
non-equilibrium state of incoming electrons, while 
the time-dependent phase $\delta(t)$ encodes the information about
the interaction. For a LL adiabaticaly connected to
reservoirs the phase is \cite{GGM_long2010}
\begin{equation}
 \delta(t) = 
\delta_1 \Theta (|\tau|/2-t)\sign\tau
%\equiv\delta_1 w_\tau(t)
\,,
\end{equation}
 where $\Theta(t)$ is the Heaviside step function. 
This allows one to reduce the operator in Eq.~(\ref{DetDef})
%(upon discretization and regularization) 
to a Toeplitz matrix \cite{Gutman10}, with
$1+n(\epsilon)(e^{-i\delta_1}-1)$ determining its symbol. (For a non-adiabatic
coupling one gets
a sequence of rectangular pulses in $\delta(t)$, yielding in the long-wire limit
a product of Toeplitz determinants.) When
$n(\epsilon)$  has discontinuities  
(``Fermi edges''), the  single particle GF acquires  a non-trivial power-law
behavior. This is in particular the case for multi-step distributions 
\begin{equation}
\label{multistep-distribution}
n(\epsilon) = \left\{
\begin{array}{ll}
1 \equiv a_0 \,, & \qquad \epsilon < \epsilon_0 \\
a_1\,, & \qquad \epsilon_0 < \epsilon <\epsilon_1 \\
\ldots &  \\
 a_m\,, & \qquad \epsilon_{m-1} < \epsilon <\epsilon_m \\
0 \equiv a_{m+1}\,, & \qquad \epsilon_m < \epsilon\,. 
\end{array}
\right.
\end{equation}
The low-energy behavior of the single-particle correlation functions
can be understood with the help of the generalized version
\cite{Gutman10,Protopopov2012}  of the
Fisher-Hartwig conjecture \cite{deift09} (see also \cite{Ivanov2011, Kozlowski2008}).
Under generic  conditions, it yields a power-law energy dependence  masked by
dephasing \cite{Gutman10,Protopopov2012}. When the electronic system is in a
pure state, i.e. for (\ref{multistep-distribution}) with all $a_i$ equal $0$ or
$1$, the dephasing is absent and the effects of correlations are
particularly strongly pronounced.

Higher-order GF of a non-equilibrium LL can be
cast in the  form  Eq.(\ref{DetDef}) as well, which offers a remarkable
opportunity of exploring  many-particle correlations in a quantum many-body 
system out of equilibrium. However, there is a serious complication as compared
to the single-particle GF case.
%The difficulty is in the fact that the
%corresponding matrices are {\it not} of the Toeplitz type anymore.
%Indeed, 
Since each creation or annihilation of an electron induces a jump in 
$\delta(t)$, the latter has a form
\begin{equation}
\label{multipulse-delta}
\delta(t) = \left\{
\begin{array}{ll}
0 \equiv \delta_0 \,, & \qquad t < t_0 \\
\delta_1\,, & \qquad t_0 < t <t_1 \\
\ldots &  \\
 \delta_k\,, & \qquad t_{k-1} < t <t_k \\
0 \equiv \delta_{k+1}\,, & \qquad t_k < t\,.  
\end{array}
\right.
\end{equation}
As we see, the determinant is now generated  by functions $A(t)$ and
$B(\epsilon)$ that both have multiple jumps and
therefore is {\it not} of Toeplitz type. Since time and
energy enter Eq.~(\ref{DetDef}) on equal footing, 
one may expect the $t$ and $\epsilon$ discontinuities to play a similar role in
the asymptotic behaviour. This suggests that there should exist  a 
generalization
of the Fisher-Hartwig formula valid for this class of determinants. This
generalization represents one of key results of the present paper.

%%%%%%%%%%%%%%%%%%%%%%%%%%%%%%%%%%%%%%%%%%%%%%%%%%%%%%%%%%%%%%%%%%%%%%%%%%%%%%%%
\begin{figure}
 \includegraphics[width=230pt]{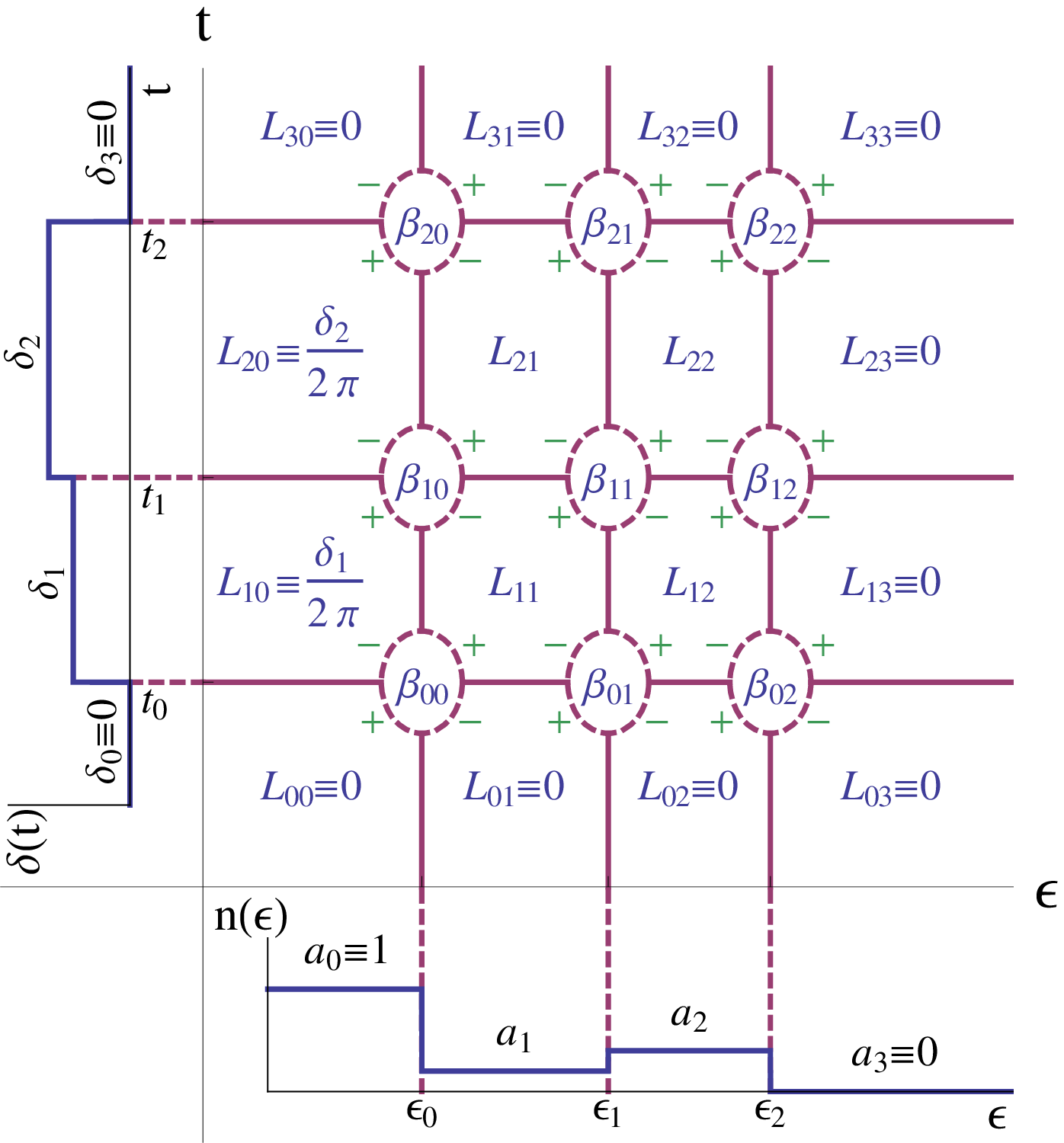}
\caption{\small Construction of the matrix $\beta_{ij}$, Eq.~(\ref{beta_ij}),
determining the
asymptotic behavior of the Fredholm determinant. } 
\label{MainPlot}
\end{figure}
%%%%%%%%%%%%%%%%%%%%%%%%%%%%%%%%%%%%%%%%%%%%%%%%%%%%%%%%%%%%%%%%%%%%%%%%%%%%%%%

We first state our result for the long-time behavior of the determinant
(\ref{DetDef}) and then present arguments in favor of it.
To formulate the result, it is convenient to draw a grid in the time-energy
plane defined by points where $A(t)$ and $B(\epsilon)$ exhibit jumps,
as shown in  Fig.~\ref{MainPlot} (for the case $k=2, m=2$). This divides the
plane in $(k+2)\times(m+2)$ rectangles (with the ``external'' ones
extending to infinity). With each of them we associate a number
\begin{eqnarray}
&& L_{ij}= (i/2\pi)\log\left[1+(e^{-i\delta_i}-1)a_j\right]; \nonumber \\
&& \hspace*{2cm} 0\leq i\leq k+1\,, \ \ 0\leq j\leq m+1.
 \label{L-lp}
\end{eqnarray}
We impose the condition that a branch of the logarithm at infinity is fixed by
$L_{i0}\equiv\delta_i/2\pi$ and  $L_{i, m+1}=L_{0j}=L_{k+1, j}\equiv0$,   while
for finite pieces any branch can be chosen.  
We now define a matrix 
\begin{equation}
\beta_{ij}=L_{i, j}+L_{i+1,j+1}-L_{i+1, j}-L_{i, j+1}\,,
\label{beta_ij}
\end{equation}
where indices $i=0,\ldots, k$ and $j=0,\ldots m$ correspond to  the grid
lines \cite{CommentLocalLog}, and a set of exponents 
\begin{equation}
 \gamma_{ii';jj'}=-\beta_{ij}\beta_{i'j'}-\beta_{ij'}\beta_{i'j}\,.
\end{equation}
The  asymptotic behaviour of  the normalized (to its zero-temperature
form) determinant
$\overline{\Delta}[\delta(t), n(\epsilon)]\equiv\Delta[\delta(t),
n(\epsilon)]/\Delta[\delta(t), T=0]$ is given by
%\begin{widetext}
\begin{eqnarray}
 \overline{\Delta}[\delta(t), n(\epsilon)] &=& \sum_{\rm br.}
C[\beta] \exp \big(-i\sum_{i\,,j}t_i\beta_{ij}\epsilon_j\big)
\nonumber \\
&\times& \prod_{\substack{i,i'=0\\\,i'<i}}^{k}\;\prod_{\substack{j,j'=0\\\,j'<j}}^{m}
\left [ (t_i-t_{i'})(\epsilon_j-\epsilon_{j'})\right]^{\gamma_{ii';jj'}}\!\!.\quad
\label{MainRes}
\end{eqnarray}
%\end{widetext}
Here the sum $\sum_{\rm br.}$ runs over all possible branches of the
logarithms in (\ref{L-lp}), and $C[\beta]$ are numerical coefficients
\cite{RemarkGfunctions}. Equation (\ref{MainRes}) represents the infrared
asymptotics which holds provided that  $(t_i-t_i')(\epsilon_j-\epsilon_j')\gg1$
for all $i > i'$ and $j > j'$. If for some set $(i,i',j,j')$
the opposite inequality holds, the corresponding factor should
be dropped in (\ref{MainRes}). When all inequalities
$(t_i-t_{i'})(\epsilon_j-\epsilon_{j'})\gg1$ are fulfilled, one can calculate
the total power of each of the factors $t_i-t_{i'}$ and $\epsilon_j-\epsilon_{j'}$ in
(\ref{MainRes}) by using the sum rules $\sum_i\beta_{ij}=0$ and
$\sum_j\beta_{ij} = (\delta_{i+1}-\delta_i)/2\pi$, which yields
%the result is given by
%Eq.~(47) of Supporting Material \cite{supplementary}. 
\begin{eqnarray}
\Delta &=& \sum_{\rm br.}
C[\beta] \exp \big[-i\sum_{i\,,j}t_i\beta_{ij}(\epsilon_j-\Lambda)\big]
\nonumber \\
&\times& \prod_{i'<i} [\Lambda(t_i-t_{i'})]^{\mu_{ii'}}
\prod_{j'<j} [(\epsilon_j-\epsilon_{j'})/\Lambda]^{\nu_{jj'}},
\label{MainRes-2}
\end{eqnarray}
where $\mu_{ii'} = \sum_j \beta_{ij} \beta_{i'j}$, $\nu_{jj'} = \sum_i \beta_{ij}
\beta_{ij'}$, and 
$\Lambda$ is the ultraviolet cutoff.

While we have no mathematical proof of Eq.~(\ref{MainRes}), we have strong
evidence in favor of its validity. Heuristically, Eq.~(\ref{MainRes})
represents a natural extension of the generalized Fisher-Hartwig formula
of Refs.~\cite{Gutman10,Protopopov2012} (valid for Toeplitz determinants) onto
the present case. Indeed, the
Fisher-Hartwig formula suggests that power-law factors in the asymptotics of
Toeplitz determinants are due to the presence of singular points $(t_i,
\epsilon_j)$ and each pair of such points contributes independently to the
result. Physically, the contribution of the pair of points $(t_i, \epsilon_j)$
and $(t_{i'}, \epsilon_{j'})$ represents the effect of particle-hole pair
scattering from the vicinity of the Fermi edge 
$\epsilon_j$ to Fermi edge $\epsilon_{j'}$ within the time $t_i-t_{i'}$. 
Combined with the symmetry between $\epsilon$ and $t$, these arguments naturally
lead to Eq. (\ref{MainRes}). On a technical level, we have proven a
generalization of strong Szeg\H{o} limit
theorem \cite{Szego52, Widom76} to the case of a multistep $A(t)$ and smooth
$B(\epsilon)$. 
Substituting in this formula a multi-step $B(\epsilon)$, we obtain logarithmic
divergencies in the exponent, yielding power laws of Eq.~(\ref{MainRes}). We
refer the reader to Supplementary Material \cite{supplementary} for detail.
Finally, we evaluated the determinant numerically for the simplest non-Toeplitz
case, with $\delta(t)$ having three jumps. The results \cite{supplementary}
confirm Eq.~(\ref{MainRes}).

%%%%%%%%%%%%%%%%%%%%%%%%%%%%%%%%%%%%%%%%%%%%%%%%%%%%%%%%%%%%%%%%%%%%%%%%%%
\begin{figure}
 \includegraphics[width=110pt]{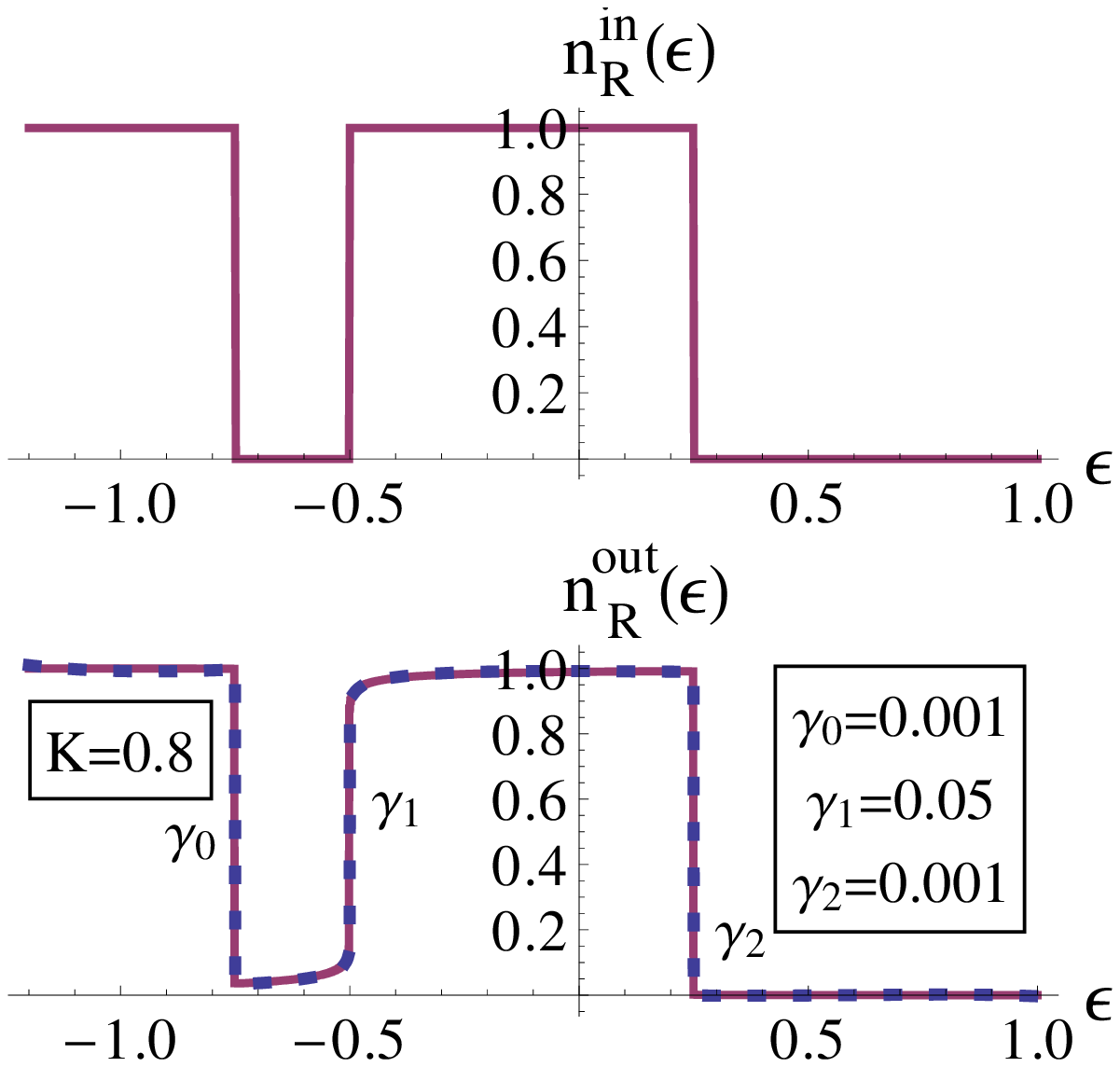}\qquad
\includegraphics[width=110pt]{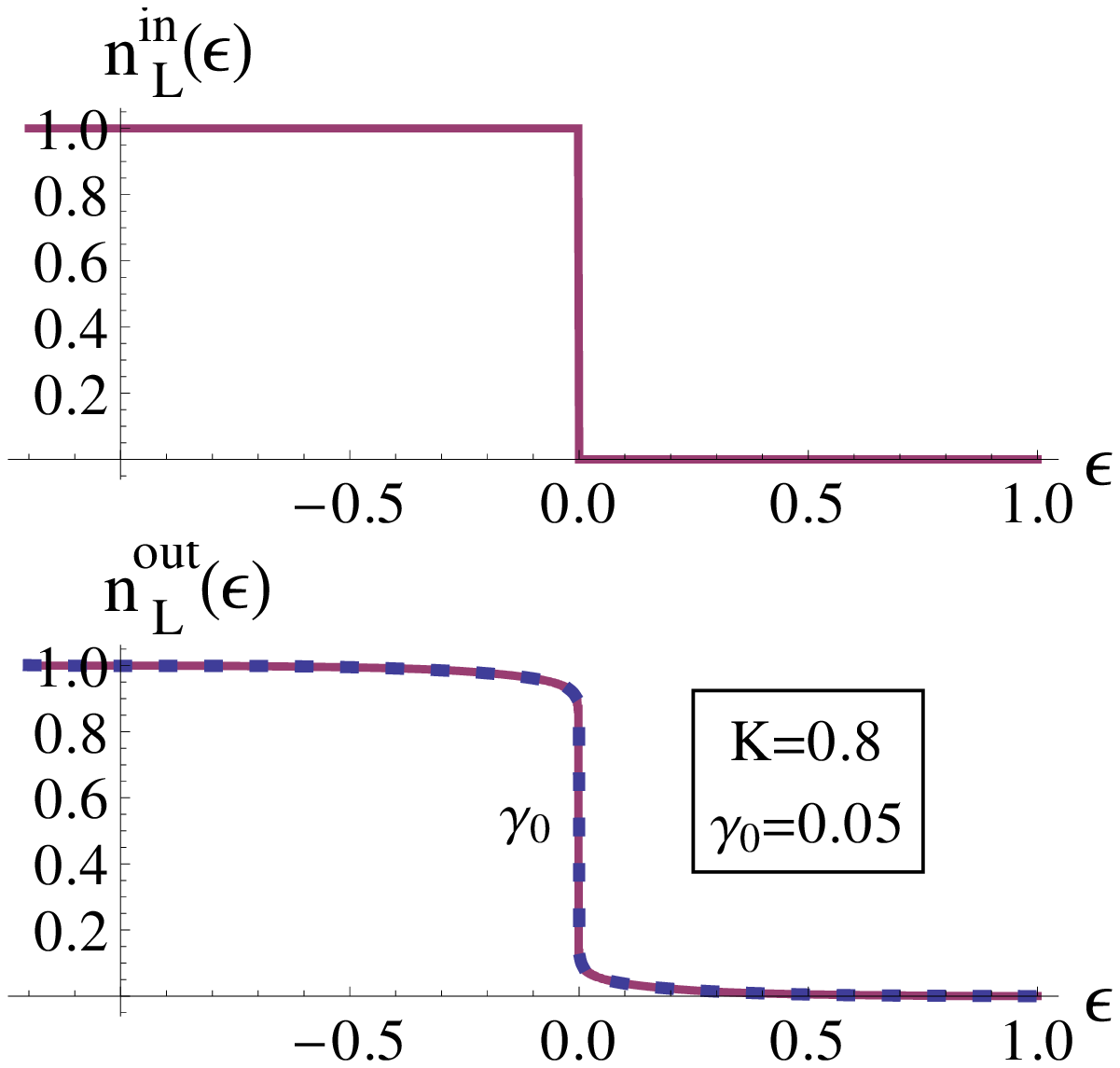}
\caption{\small Distribution functions of the incoming (top) and outgoing
(bottom) electrons. The incoming right-mover distribution function has
three Fermi edges and exhibits inverse population. Incoming left movers 
are assumed to have zero-temperature Fermi-Dirac distribution.
Distribution functions of  fermions at the
output of the device show power-law singularities at Fermi edges with exponents
indicated in the legends.  
To generate the plot we assumed $K=0.8$}
\label{Distributions}
\end{figure}
%%%%%%%%%%%%%%%%%%%%%%%%%%%%%%%%%%%%%%%%%%%%%%%%%%%%%%%%%%%%%%%%%%%%%%%%%%%%

%%%%%%%%%%%%%%%%%%%%%%%%%%%%%%%%%%%%%%%%%%%%%%%%%%%%%%%%%%%%%%%%%%%%%%%%%%%%
\begin{figure}
 \includegraphics[width=180pt]{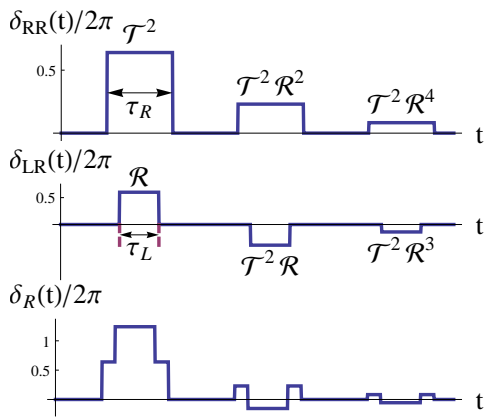}\qquad
\caption{\small {\it Top and Middle}. Phases $\delta_{RR}(t)$ and
$\delta_{RL}(t)$ controlling the distributions of left- and right- mover at the
output of the wire. The time interval  $2Kl/v_F$ between different pulses is
macroscopically  
large  and the corresponding determinants can be approximated as products. {\it
Bottom}. The phase $\delta_R(t)= 
\delta_{RR}(t)+\delta_{RL}(t)$ controlling the correlator $\langle\langle 
n_R(x_R, \epsilon_R)n_L(x_L, \epsilon_L)\rangle\rangle$ at $x_R+x_L=-Kl$. } 
\label{Phases}
\end{figure}
%%%%%%%%%%%%%%%%%%%%%%%%%%%%%%%%%%%%%%%%%%%%%%%%%%%%%%%%%%%%%%%%%%%%%%%%%%%

%%%%%%%%%%%%%%%%%%%%%%%%%%%%%%%%%%%%%%%%%%%%%%%%%%%%%%%%%%%%%%%%%%%%%%%%%%%%%%
\begin{figure*}
 \includegraphics[width=470pt]{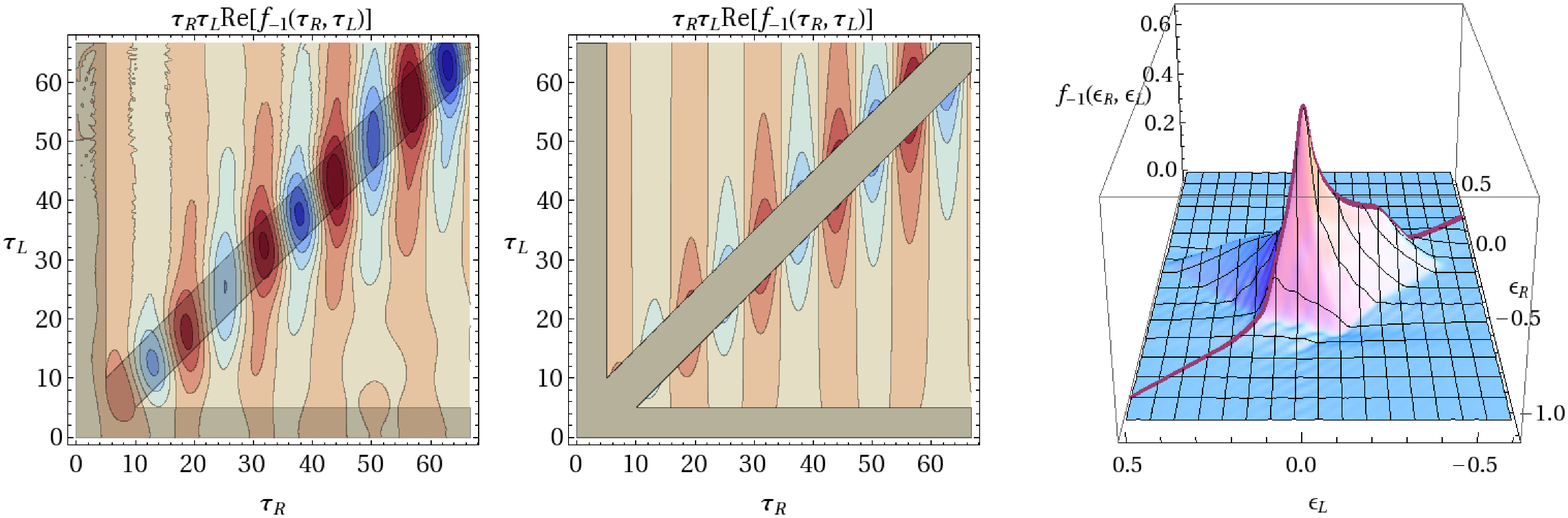}
\caption{\small {\it Left and Middle.} Contour plot of the irreducible
correlation function  
$f_{-1}(\tau_R, \tau_L)$   multiplied by $\tau_R\tau_L$. The left panel is
the result of numerical
evaluation of Fredholm determinants. The middle panel shows the fit based on
Eqs.(\ref{delta1}) and (\ref{delta2}) with coefficients $C_i$ used as fitting
parameters. Shadowed regions on both graphs are outside the applicability of
the asymptotic expansion (\ref{MainRes}). 
{\it Right.} Irreducible correlator of fermionic occupation numbers  
in energy representation. 
 }
\label{RLCorrelator}
\end{figure*}
%%%%%%%%%%%%%%%%%%%%%%%%%%%%%%%%%%%%%%%%%%%%%%%%%%%%%%%%%%%%%%%%%%%%%%%%%%

We now apply Eq.(\ref{MainRes})  to the  problem of correlations in a
non-equilibrium LL.   
We consider a LL conductor of length $l$ (characterized by LL parameter $K$)
driven out of equilibrium via the injection of electrons 
with distributions $n^{\rm in}_\eta (\epsilon)$  [$\eta=R$ ($L$) for 
right- (left-) movers] from the non-interacting leads, see Fig. 1 of
Ref.\cite{Protopopov2011}. 
To be specific, we assume that the right-moving electrons have the triple-step
distribution function with population inversion (Fig. \ref{Distributions}) while
the left movers are at zero temperature. This is the simplest non-equilibrium
setup without dephasing.  The width 
$\epsilon_2-\epsilon_0$ of $n^{\rm in}_R(\epsilon)$ is set to unity, thus
determining the time and energy scale of the problem.  
We model the leads by 1D conductors with $K=0$.
The correlations effects discussed in this paper can be traced back to  the
scattering of  LL plasmons at the boundaries of the wire. 
This  scattering is strong when boudaries are sharp (compared to the plasmon
wave length),  $K(x)=K\Theta\left(l/2-|x|\right)$, so that we focus on this
regime. We stress that we assume the absence of
the {\it electron} backscattering in the wire, i.e.,  $K(x)$ is smooth on the
scale of the Fermi wavelength.

At thermal equilibrium  ($n^{\rm in}_{R,L}(\epsilon)$ are  Fermi-Dirac
distributions with equal temperatures), the 
interaction causes no correlations for electrons outside the LL wire. 
In other words, electrons leave  the  wire in the same state as free fermions
would do
at given temeparture. The situation changes dramatically under non-equilibrium
conditions \cite{Protopopov2011}. 
The plasmon scattering at the boundaries of the wire 
(characterized by reflection coefficient ${\cal R}=(1-K)/(1+K)$ and transmission
coefficient ${\cal T}=\sqrt{1-{\cal R}^2}$) not only leads to an electron energy
redistribution but also induces correlations between outgoing electrons.
As we found earlier \cite{Protopopov2011},   distribution functions 
$n_{R,L}^{\rm out}(\tau)$ in the non-interacting regions are  given (up
to a numerical factor) by the determinant (\ref{DetDef})
with phases $\delta_{RR}(t)$ and $\delta_{LR}(t)$ shown in Fig.~\ref{Phases}. 
Both phases consist of an infinite sequence of rectangular pulses
separated by a time interval $2Kl/v_F$. 
In the long-wire limit the corresponding determinant  can be decomposed into the
product of Toeplitz determinants controlled by individual
pulses \cite{footnote2}. An asymptotic analysis of these determinants
\cite{Protopopov2012} leads to the conclusion that  distribution functions of
outgoing electrons have power-law singularities at the Fermi edges
(Fig.~\ref{Distributions}). Note that for relatively weak interaction  the
interaction effects are most pronounced at the inverted edge of the
$n_R(\epsilon)$ signaling that the dominant physical process is the scattering
of electron-hole pairs from this edge to the Fermi edge of $n_L(\epsilon)$.

To reveal the correlations induced by the plasmon scattering, we 
consider  the irreducible two-particle distribution function of outgoing
electrons in the two leads,
$f(\epsilon_R, \epsilon_L, x_R+x_L)=\langle\langle \hat{n}_R(x_R,
\epsilon_R)\hat{n}_L(x_L, \epsilon_L)\rangle\rangle$ with $x_R>l/2$ and
$x_L<-l/2$. It is a non-trivial function of $x_R+x_L$, with the correlations
being most pronounced in a vicinity of the points  $x_R+x_L=(2m+1)Kl$ 
\cite{Protopopov2011}. We focus on the case
$x_R+x_L=-Kl$ where the correlations are maximal. At this point the function
$f(\epsilon_R, \epsilon_L, -Kl)\equiv f_{-1}(\epsilon_R, \epsilon_L)$ is given
by the determinant (\ref{DetDef}) with the phase
$\delta_R(t)=\delta_{RR}(t)+\delta_{LR}(t)$, see Fig. \ref{Phases}.   
 Explicitly
%\begin{widetext}
\begin{eqnarray}
\hspace*{-0.5cm} f_{-1}(\epsilon_R, \epsilon_L) \!\! &=& \!\! 
-\frac{1}{4\pi^2}\int \frac{d\tau_R d\tau_L}{\tau_R \tau_L} 
e^{i\epsilon_R\tau_R+i\epsilon_L\tau_L} \nonumber \\
\!\! & \times& \!\! \left(\overline{\Delta}[\delta_{R}, n_R^{\rm in}] -
\overline{\Delta}[\delta_{RR}, n_R^{\rm in}]\overline{\Delta}[\delta_{RL},
n_R^{\rm in}]\right)\!.
\end{eqnarray}
 %\end{widetext}
Investigation of the correlations of left- and right- movers is now reduced to
calculation  of the Fredholm determinant  
with phase $\delta_R(t)$ that  can be  readily done by employing  
Eq.~(\ref{MainRes}). 
For a long wire the determinant $\overline{\Delta}[\delta_R, n_R^{\rm in}]$ 
decomposes into a product of determinants corresponding to  individual pulses
forming $\delta_R(t)$ (see Fig.~\ref{Phases}).
We write $\overline{\Delta}[\delta_R, 
n_R^{\rm in}]=\overline{\Delta}_1\overline{\Delta}_2$, where
$\overline{\Delta}_1$ is a contribution of the first pulse and
$\overline{\Delta}_2$ of the remaining pulses.
For definiteness, we focus on the case of a weak interaction (small
${\cal R}$) when the phase $\delta_R(t)$ is small outside the
first pulse. Accordingly,  
the asymptotic behavior of the determinants for all pulses but the first
one is governed by a single term in the sum (\ref{MainRes}) corresponding to a
choice of logarithm branches such that  $|L_{ij}|<\pi$ for all $i$ and $j$. 
 [The analysis for an arbitrary interaction proceeds in the same
way; one just may need to take into account several further terms in
Eq.~(\ref{MainRes}).]
This yields
\begin{equation}
 \overline{\Delta}_2 = C_0
\left(\frac{\tau_R+\tau_L}{\tau_R-\tau_L}\right)^{\frac{4{\cal T}^2{\cal
R}^3}{1+{\cal R}^2}}
\tau_R^{-\frac{2{\cal T}^2{\cal R}^4}{1+{\cal R}^2}}\tau_L^{-\frac{2{\cal
T}^2{\cal R}^2}{1+{\cal R}^2}}\,.
\label{delta1}
\end{equation}
To find $\overline{\Delta}_1$, we note that   $\delta_R(t)$ is
close to $2\pi$; an inspection of Eq.~(\ref{MainRes}) leads 
to the following three dominant contributions:
%\begin{widetext}
\begin{eqnarray}
 \overline{\Delta}_1 &=& \left(C_1 e^{-i\epsilon_0 \tau_R}+C_2
e^{-i\epsilon_2 \tau_R}\right)
\left(\frac{\tau_R+\tau_L}{\tau_R-\tau_L}\right)^{4{\cal R}^3} \nonumber \\
&\times& \tau_R^{-2{\cal R}^4}\tau_L^{-2{\cal R}^2}+C_3e^{-i\epsilon_1
\tau_R}\left(\frac{\tau_R+\tau_L}{\tau_R-\tau_L}\right)^{4{\cal R}\left({\cal
R}^2+1\right)} \nonumber\\
&\times& \tau_R^{-2{\cal R}^2\left({\cal R}^2+2\right)}\tau_L^{-2{\cal R}^2}\,.
\label{delta2}
\end{eqnarray}
% \end{widetext}
Since we are not interested in the dependence on energies 
$\epsilon_i-\epsilon_j$, we have absorbed it in the coefficients $C_i$ in
Eqs.~(\ref{delta1}) and (\ref{delta2}). Combining (\ref{delta1}) and
(\ref{delta2}) with asymptotic expansion for the
Toeplitz determinants 
$\overline{\Delta}[\delta_{RR}]$ and  $\overline{\Delta}[\delta_{RL}]$
\cite{supplementary}, we arrive at the
asymptotic expansion of irreducible correlation function $f_{-1}$ in time
domain. 

The two-particle correlation function $f_{-1}(\tau_R, \tau_L)$ is shown in Fig.
\ref{RLCorrelator} for $K=0.8$. The left panel
presents the results of direct numerical evaluation of the Fredholm
determinants, while the middle panel displays the asymptotics given by Eqs.
(\ref{delta1}) and (\ref{delta2}) with the coefficients $C_i$ used as fitting
parameters.  As expected, the asymptotics based on Eq.~(\ref{MainRes})
reproduces correctly the behavior of $f_{-1}(\tau_R, \tau_L)$
outside the region of $(\tau_R, \tau_L)$-plane where $\tau_R\lesssim 1$ or
$\tau_L\lesssim 1$,  or $|\tau_R-\tau_L|\lesssim 1$. 

Power-law long-time behavior translates into singularities 
in the energy reperentation of the  correlation function
 $f_{-1}(\epsilon_R, \epsilon_L)$ plotted in  right panel of
Fig.\ref{RLCorrelator}. 
%Physically, it  is natural  to characterize the correlations of fermionic
%occupation numbers in energy domain. Left panel of Fig.%\ref{RLCorrelator}
%demonstrates the result of numerical evaluation of $f_{-1}(\epsilon_R,
%\epsilon_L)$.  
%The resulting behavior of $f_{-1}(\epsilon_R, \epsilon_L)$ is quite complex. 
Strong correlations are observed that form a ``crest'' going along the line
$\epsilon_R+\epsilon_L=\epsilon_1$ (thick red line). The analysis shows that 
the main contribution to this structure in $f_{-1}(\epsilon_R, \epsilon_L)$
comes from the vicinity of the main diagonal  $|\tau_R-\tau_L|\lesssim 1$ of
the $(\tau_R,\tau_L)$-plane. In this region, the main
contribution to Eq.~(\ref{delta2}) is given by the last term yielding
$\overline{\Delta}_1 \sim e^{-i\epsilon_1
\tau_R}(\tau_R+\tau_L)^{\alpha_1}$ with $\alpha_1 = 4{\cal R} - 6{\cal R}^2 + 4
{\cal R}^3 - 2{\cal R}^4 $, whereas Eq.~(\ref{delta1}) reduces to 
$\overline{\Delta}_2 \sim (\tau_R+\tau_L)^{\alpha_2}$ with $\alpha_2 = - 2
{\cal T}^2 {\cal R}^2 (1 - {\cal R})^2  / (1 + {\cal R}^2)$. 
%
%The emergence of this crest can be traced back to the last
%term in Eq.~(\ref{delta2}). It turns out that  
%significant contribution to $f_{-1}(\epsilon_R, \epsilon_L)$ comes from the
%vicinity of the main diagonal  
%$|\tau_R-\tau_L|\lesssim 1/(\epsilon_2-\epsilon_0)$ where the last term in
%Eq.(\ref{delta2}) reduces to 
%\begin{equation}
% e^{-i\epsilon_1 \tau_R}\left(\tau_R+\tau_L\right)^{4{\cal R}\left({\cal
%R}^2+1\right)}
%\tau_R^{-2{\cal R}^2\left({\cal R}^2+2\right)}\tau_L^{-2{\cal R}^2}\,.
%\end{equation}
%As expected  power-low decay demonstrated by asymptotic (\ref{delta2}) gets translated into power-low singularities 
%of $f_{-1}(\epsilon_R, \epsilon_L)$. 
%%The main feature seen on Fig.\ref{RLCorrelator} is a crest going along the line $\epsilon_R+\epsilon_L=\epsilon_1$
%%marked by solid line.  Its appearance can be traced back to the last term in Eq. (\ref{delta2}). It turns out that 
%%significant contribution to $f_{-1}(\epsilon_R, \epsilon_L)$ comes from the vicinity of the main diagonal 
%%$|\tau_R-\tau_L|\lesssim 1/(\epsilon_2-\epsilon_0)$ where the last term in Eq.(\ref{delta2}) reduces to
%
%Combining this with the appropriate reduction of $\overline{\Delta}_1$ 
%
This yields the behavior of $f_{-1}(\epsilon_R, \epsilon_L)$ near the crest
\begin{equation}
 f_{-1}(\epsilon_R,
\epsilon_L)\sim
|\epsilon_R+\epsilon_L-\epsilon_1| ^ { 1- \alpha_1 - \alpha_2}\,.
\label{two-particle-FES}
\end{equation}
Physically, the crest  originates from intensive scattering of
electron-hole pairs from the vicinity of the left-mover Fermi surface $\epsilon
= 0$ to the inverted Fermi surface $\epsilon_1$ of right movers, and
Eq.~(\ref{two-particle-FES}) can be viewed as a Fermi-edge singularity in a
higher-order correlation function.   

To summarize, we have studied correlation induced by interaction in a
non-equilibrium LL. We have shown that electron scattering between multiple
Fermi edges leads to power-law singularities in  
many-particle distribution functions. As a particular example, we calculated a
two-particle  distribution
function characterizing correlations between left- and right-moving outgoing
fermions. Technically, many-particle correlation
functions are expressed in terms of Fredholm determinants ${\rm
det}(1+\hat{A}\hat{B})$, where $A(\epsilon)$ and $B(t)$ have multiple
discontinuities in energy and time spaces, respectively.
We have conjectured a general asymptotic formula for this class of
determinants and provided an ample analytical and numerical support to this
conjecture. The results are expected to be relevant to a broad class of
non-equilibrium many-body problems including the Kondo problem; work in this
direction is currently underway.

We acknowledge support by Alexander von
Humboldt Foundation, ISF, and GIF.

\begin{widetext}
\appendix
\section{Analytical justification of the asymptotic formula for singular
Fredholm determinants}

In this section we present analytical arguments in favor of the 
Eqs. (8), (9) of the main text for the asymptotic behavior of the determinants 
${\rm det}(1+\hat{A}\hat{B})$ with $A(\epsilon)$ and $B(t)$ having multiple
discontinuities in energy and time spaces, respectively. We
begin [Sec.~\ref{S:SzegoVSFisherHartwig}] by reminding the reader about 
Szeg\H{o} formula for Toeplitz matrices with smooth symbols and a
generalized version of the Fisher-Hartwig formula for Toeplitz matrices with
singular symbols. We emphasize there a connection between these two formulas. 
We proceed then by formulating and presenting a proof of a generalization of
Szeg\H{o}  formula for the case of a multi-step
function $A(t)$ and smooth $B(\epsilon)$, Sec.~\ref{subsec:gen-Szego}.
Finally, in Sec.~\ref{subsec:gen-Fisher-Hartwig} we use this result to obtain
the scaling behavior of Fredholm determinants in the case when both functions 
$A(t)$ and $B(\epsilon)$ have multiple singularitites. 
 
%explore the interrelation between the strong Szeg\H{o} limit theorem and
%classical Fisher-Hartwig conjecture. We also show that appropriate
%generalization of the strong Szeg\H{o} formula to the case  of multi-step 
%counting phase provides heuristic justification  to our generalization of
%Fisher-Hartwig conjecture.

\subsection{Toepliz determinants:\ \ Strong Szeg\H{o} limit theorem and
generalized Fisher-Hartwig conjecture}
\label{S:SzegoVSFisherHartwig}

Let us consider the determinant of $N\times N$ Toeplitz matrix $T_N[f]=\left\{f_{i-j}\right\}$, $0\leq i, j\leq N-1$ generated by function $f(z)$ defined on the unit circle $z=e^{i\theta}$ and having sufficiently smooth logarithm $V(z)\equiv \ln f(z)$. We use the notation
\begin{equation}
 f_{i-j}=\oint f(z) z^{-(i-j)} \frac{dz}{2\pi i z}\,.
\end{equation}
Smoothness of $V(z)$ on the unit circle implies  the existence of the Wiener-Hopf decomposition for $f$
\begin{equation}
 f(z)=e^{V_+(z)}e^{V_0}e^{V_-(z)}\,, \qquad V_0=\frac{1}{2\pi i}\oint f(z)\frac{dz}{z}
\end{equation}
with functions $V_{\pm}(z)$ being analytic inside and outside the unit circle respectively. In terms of Fourier components of $V(z)$
\begin{equation}
 V_+(z)=\sum_{k=1}^\infty V_k z^k\,, \qquad V_-(z)=\sum_{k=1}^{\infty} V_{-k}z^{-k}\,.
\end{equation}
The strong Szeg\H{o} limit theorem~\cite{Szego52} states  that under the  assumptions made above  the asymptotic behavior of the Toeplitz determinant $\det T_N[f]$ at $N\rightarrow\infty$ is given by
\begin{equation}
 \det T_N[f]=e^{NV_0}\left(\exp\left[\sum_{k=1}^{\infty}kV_k V_{-k}\right]+\smallO{1}\right)\,.
\label{SzegoClassical}
\end{equation}
When interpreted in physical terms Eq.(\ref{SzegoClassical}) yields the
long-time asymptotic of the single-particle correlation functions of $1D$
interacting fermions in a non-equilibrium state characterized by {\it smooth}
distribution function \cite{Gutman10}. The exponential term in
(\ref{SzegoClassical}) encodes herewith the information about the oscillations
of the Green functions and its exponential decay due to non-equilibrium
dephasing, while the precise value of the pre-exponential factor is of little
importance.  

The situation changes significantly if one considers single-particle Green functions in a state characterized by distribution function of electrons having $m+1=1, 2,\ldots$ discontinuities (Fermi edges). 
To be specific we limit our discussion to the case of multi-step distribution  function 
\begin{equation}
\label{multistep-distribution_app}
n(\epsilon) = \left\{
\begin{array}{ll}
1 \equiv a_0 \,, & \qquad \epsilon < \epsilon_0 \\
a_1\,, & \qquad \epsilon_0 < \epsilon <\epsilon_1 \\
\ldots &  \\
 a_m\,, & \qquad \epsilon_{m-1} < \epsilon <\epsilon_m \\
0 \equiv a_{m+1}\,, & \qquad \epsilon_m < \epsilon\,,  
\end{array}
\right.
\end{equation}

Supplementing the theory with the ultraviolet cutoff $\Lambda$, one ends up with
a Toeplitz matrix with symbol  having  $m+1$ jumps. In the context
of Toeplitz matrices such discontinuitites are known as Fisher-Hartwig
singularities. More specifically, the symbol of interest is 
\begin{eqnarray}
 f(z)=e^{V_0}z^{\sum_{j=0}^{m}\beta_j}\prod_{j=0}^{m}g_{z_j, \beta_j}(z)z_j^{-\beta_j}\,,
\label{FisherHartwigSymbol}
\\
V_0=-i\pi\sum_j\beta_j+\sum_j\beta_j\ln z_j\,,
\label{V0}
\end{eqnarray}
 with $z_j= e^{i\pi\epsilon_j/\Lambda}\equiv e^{i\theta_j} $ and 
\begin{equation}
 g_{z_j, \beta_j}(z)=\left\{
\begin{array}{c}
 e^{i\pi\beta_j}\,, \qquad -\pi <\arg z<\theta_j\\
 e^{-i\pi\beta_j}\,, \qquad \theta_j<\arg z <\pi\,.
\end{array}
\right.
\end{equation}
Here the numbers $\beta_j$ controlling the jumps of $V(z)\equiv\ln f(z)$ at
points $z_j$ are given by
\begin{equation}
\beta_j=\frac{1}{2\pi i}\ln \frac{1+(e^{-i\delta}-1)a_j}{1+(e^{-i\delta}-1)a_{j+1}}\\
=\frac{1}{2\pi i}\left[V(\theta_j-0)-V(\theta_j+0)\right]\,.
\end{equation}
We are interested in the large-$N$ asymptotic behavior of the determinant $\det
T_N[f]$. In physical language the size of the Toeplitz matrix corresponds to the
time in the correlation function $\tau=\pi N/\Lambda$. 

%%%%%%%%%%%%%%%%%%%%%%%%%%%%%%%%%%%%%%%%%%%%%%%%%%%%%%%%%%%%%%%%%%%%%%%%%%%%%%%
\begin{figure}
 \includegraphics[width=220pt]{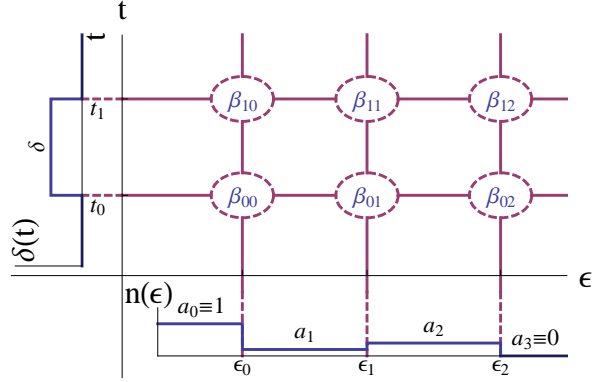}
\caption{\small Pictorial representation of Eq.~(\ref{SzegoCrude}). A set of
numbers $\beta_{li}$ is associated with the set of crossing points
$(\epsilon_i, t_l)$ of singularities in $(\epsilon, t)$
plane. Each pair of such crossing points with different time coordinates gives
rise to a logarithmically diverging contribution to the exponent in
Eq.~(\ref{SzegoCrude}). }
\label{BettaGraphSimplest}
\end{figure}
%%%%%%%%%%%%%%%%%%%%%%%%%%%%%%%%%%%%%%%%%%%%%%%%%%%%%%%%%%%%%%%%%%%%%%%%%%%%%%

Strictly speaking, the strong Szeg\H{o} theorem does not apply to the
Toeplitz matrix with a singular symbol (\ref{FisherHartwigSymbol}); the
corresponding extension of the large-$N$ limit theorem is known as
Fisher-Hartwig conjecture \cite{FisherHartwig}.
This conjecture states that at large $N$
\begin{equation}
 \det T_N[f]=C[\beta]e^{NV_0} N^{-\sum_{j=0}^{m}\beta_j^2}\prod_{0\leq j<i\leq
m}|z_i-z_j|^{2\beta_i\beta_j}\,,
\label{FisherHartwigResultN}
\end{equation}
with $C[\beta]$ being a (known) numerical coefficient. Casting  this  result in
terms of  physical variables and assuming that all energies $\epsilon_i$ are
small compared to $\Lambda$, one obtains
\begin{equation}
 \det
T_N[f]=C[\beta]e^{\frac{i\delta\tau\Lambda}{2\pi}+i\sum_{j=0}^m\beta_j\epsilon_j
\tau} \left(\frac{\pi}{\tau\Lambda}\right)^{\sum_{j=0}^{m}\beta_j^2}\prod_{0\leq
j<i\leq
m}\left[\frac{\pi}{\Lambda}(\epsilon_i-\epsilon_j)\right]^{2\beta_i\beta_j}\,.
\label{FisherHartwigResult}
\end{equation}

While a rigorous mathematical proof of Eq.(\ref{FisherHartwigResult}) is highly
non-trivial and for the general settings was achieved only very recently,
a strong argument in favor of Eq.(\ref{FisherHartwigResult}) can be given on
the basis of the Szeg\H{o} formula (\ref{SzegoClassical}).   Indeed, the
jumps of $V(z)$ at points $z_i=e^{i\theta_i}$ dictate the following asymptotic
behavior of the Fourier coefficients $V_k$ at $k\rightarrow\infty$ 
\begin{equation}
 V_k\sim\frac{1}{k}\sum_j\beta_j\,.
\end{equation}
The naive application of the Szeg\H{o} formula leads now to
\begin{equation}
\det T_N[f]=e^{NV_0}\exp\left[\sum_{i,j=0}^m\beta_{0i}\beta_{1j}\sum_{k=1}^{\infty}\frac{1}{k}\right],
\label{SzegoCrude}
\end{equation}
where we have introduced here $\beta_{1i}=-\beta_{0i}=\beta_i$.

Equation (\ref{SzegoCrude}) can be given a pictorial representation shown in Fig.\ref{BettaGraphSimplest}. 
We attribute the  quantities $\beta_{li}$ to the points  $(\epsilon_i, t_l)$  in
$(\epsilon, t)$-plane where jumps of the functions $n(\epsilon)$ and $\delta(t)$
cross.  Each pair of such points with different time arguments
[i.e., $(\epsilon_i, t_l)$ and $(\epsilon_{i'}, t_{l'})$ for  $t_l\neq t_{l'}$)]
provides the following contribution to the last factor in Eq.
(\ref{SzegoCrude}): 
\begin{equation}
 \exp\left[\frac12\beta_{li}\beta_{l'i'}\sum_{k=1}^{\infty}\frac{1}{k}\right]\sim
 \exp\left[\frac12\beta_{li}\beta_{l'i'}\int dt \frac{1}{t}\right]\,.
\end{equation}
 To give meaning to this expression, we have to cut off the logarithmically
diverging integral. The lower integration limit 
is set naturally by the ultraviolet cutoff $\Lambda$. For $i=i'$ one expect the
upper integration limit to be given by  
the time $\tau$. On the other hand, for $i\neq i'$ the upper limit of
integration is expected to be $1/|\epsilon_i-\epsilon_{i'}|$ (recall that the
asymptotics we are discussing is valid under condition 
$\tau|\epsilon_i-\epsilon_{i'}|\gg 1$). Using these simple cutoff rules 
together with the expression (\ref{V0}) for the coefficient $V_0$, we see that
the Szeg\H{o} theorem justifies in a natural way the
Fisher-Hartwig conjecture (\ref{FisherHartwigResult}).  

There is, however, the following subtlety here. Recalling the
definition (\ref{FisherHartwigSymbol}) of the symbol with Fisher-Hartwig
singularities, one sees that the numbers $\beta_j$ are defined modulo a set of
integer numbers $n_j$ such that 
\begin{equation}
 \sum_{j=0}^mn_j=0\,.
\label{nConstraint}
\end{equation}
In other words two sets of coefficients $\{\beta_j\}$ and
$\{\widetilde{\beta_j}=\beta_j+n_j\}$ describe the same symbol $f(z)$ and should
produce the same asymptotics. The recently proven rigorous formulation of the
Fisher-Hartwig conjecture \cite{deift09} indeed respects this requirement.
Specifically, it states that one should choose the set of 
$\beta_j$ in Eq.~(\ref{FisherHartwigResult}) in such a way that the sum 
$\sum_{j=0}^m\beta_j^2$ determining the power of $N$ takes the smallest possible
value. 

More recently, it was shown in \cite{Gutman10,Protopopov2012} that other
choices of integer shifts (i.e. of logarithm branches) in the expressions for
$\beta_j$ are also important and determine the sub-leading terms in the
asymptotic behavior of a Toeplitz determinant with a singular symbol. A
generalized Fisher-Hartwig conjecture formulated in this work reads
\begin{equation}
 \det
T_N[f]=\sum_{\{\beta\}}C[\beta]e^{\frac{i\delta\tau\Lambda}{2\pi}+i\sum_{j=0}
^m\beta_j\epsilon_j \tau}
\left(\frac{\pi}{\tau\Lambda}\right)^{\sum_{j=0}^{m}\beta_j^2}\prod_{0\leq
j<i\leq
m}\left[\frac{\pi}{\Lambda}(\epsilon_i-\epsilon_j)\right]^{2\beta_i\beta_j}
(1+\ldots)\,.
\label{FisherHartwigResultSum}
\end{equation}
The summation here goes over all choices of  $\{\beta_j\}$,
i.e., over all sets of integer shifts $n_j$ satisfying 
Eq.~(\ref{nConstraint}). Each such set
yields in Eq.~(\ref{FisherHartwigResultSum}) a term with a distinct oscillatory
exponent $e^{i\sum_{j=0}^m\beta_j\epsilon_j \tau}$.
Equation (\ref{FisherHartwigResultSum}) presents explicitly the leading
asymptotic behavior for the factor multiplying each of these exponents. Apart
from this dominant term, there will be in general also subleading (in powers of
$1/t$) terms corresponding to the same exponent; these are abbreviated
by $+\ldots$ in the last bracket.

While a rigorous proof of the generalized
Fisher-Hartwig conjecture Eq.~(\ref{FisherHartwigResultSum}) remains to be
constructed, there is little doubt that it is correct.

\subsection{Multiple singularities in $A(t)$:\ \ Generalization of Szeg\H{o}
formula}
\label{subsec:gen-Szego}

%We aim at motivating the generalization of the Fisher-Hartwig conjecture to the
%case of multi-step counting phase $\delta(t)$ presented in the main text.  It
%is instructive to work out first the variant of the strong Szeg\H{o} formula,
%appropriate for this case.  
We assume now that the electronic distribution
$n(\epsilon)$ is a smooth function of energy, whereas the scattering phases
$\delta(t)$  has  a multi-step structure: 
\begin{equation}
\label{multipulse-delta_app}
\delta(t) = \left\{
\begin{array}{ll}
0 \equiv \delta_0 \,, & \qquad t < t_0 \\
\delta_1\,, & \qquad t_0 < t <t_1 \\
\ldots &  \\
 \delta_k\,, & \qquad t_{k-1} < t <t_k \\
0 \equiv \delta_{k+1}\,, & \qquad t_k < t\,.  
\end{array}
\right.
\end{equation}
We will assume for definiteness that $k=2$; generalization to a larger
number of steps in $\delta(t)$ is completely straightforward.
Among known derivations of Szeg\H{o} formula, the operator-theory approach due
to Widom 
\cite{Widom76} permits a particularly transparent generalization to our problem.
We thus follow this approach here. 

Upon introduction of an ultraviolet cutoff $\Lambda$ the functional  determinant
of interest
\begin{equation}
 \det\left[1+\left(e^{-i\delta(t)}-1\right)n(\epsilon)\right]
\end{equation}
is reduced to the determinant of a finite matrix of the size $N\times N$ with
the structure
\begin{equation}
 A_{N_1, N_2}\left[f^{(1)}, f^{(2)}\right]=\left(
\begin{array}{c}
 T_{N_1, N}\left[f^{(1)}\right]\\
 T_{N_2, N}\left[f^{(2)}\right]
\end{array}
\right)\,, \qquad N=N_1+N_2\,.
\end{equation}
Here $T_{N_i, N}\left[f^{(i)}\right]$ stands for a rectangular Toeplitz matrix
of the size $N_i\times N $ with a symbol 
\begin{equation}
 f^{(i)}(\epsilon)=\left[1+\left(e^{-i\delta_i}-1\right)n(\epsilon)\right]e^{-i\delta_i\epsilon/2\Lambda}
\end{equation}
which is smooth on the unit circle $z=e^{i\pi\epsilon/\Lambda}$.
The numbers $N_0$ and $N_1$ are determined by $\delta(t)$ via
\begin{equation}
 N_1=(t_1-t_0)\Lambda/\pi\,, \qquad N_2=(t_2-t_1)\Lambda/\pi\,.
\end{equation}
Let us introduce a semi-infinite Hankel matrix with symbol $f$ according to
\begin{equation}
 H[f]=\left\{f_{i+j+1}\right\}\,, \qquad i, j=0,\ldots \infty
\end{equation}
and the semi-infinite matrices $P_N$ and $Q_N$ with matrix elements 
($i,j=0,\ldots \infty$)
\begin{eqnarray}
 \left(P_N\right)_{ij}=\left\{
\begin{array}{cc}
 \delta_{ij}\,, &\max(i, j)<N-1\\
  0\,, & \max(i, j)>N-1
\end{array}
\right. \ \ {\rm and} \ \ 
%\\
 \left(Q_N\right)_{ij}=\left\{
\begin{array}{cc}
 \delta_{i, N-1-j}\,, &\max(i, j)<N-1\\
  0\,, & \max(i, j)>N-1\,.
\end{array}
\right.
\end{eqnarray}

With this notations  matrix $A_{N_0, N_1}\left[f^{(0)}, f^{(1)}\right]$ can be presented in the form 
\begin{equation}
 A_{N_1, N_2}\left[f^{(1)}, f^{(2)}\right]=\left(
\begin{array}{cc}
 T_{N_1}\left[f^{(1)}\right] & Q_{N_1} H\left[\widetilde{f}^{(1)}\right]P_{N_2}\\
 P_{N_2} H\left[f^{(2)}\right]Q_{N_1} & T_{N_2}\left[f^{(2)}\right]
\end{array}
\right)\,.
\label{ABlock}
\end{equation}
Here the symbol $\widetilde{f}$ is obtained from a symbol $f$ according to
\begin{equation}
 \widetilde{f}(z)=f\left(z^{-1}\right)\,.
\end{equation}
The block representation (\ref{ABlock}) of the matrix $A_{N_0, N_1}$ allows us to write its determinant in the form
\begin{equation}
 \det A_{N_1, N_2}=\det\left[ T_{N_1}\left[f^{(1)}\right]\right]\det\left[ T_{N_2}\left[f^{(2)}\right]\right]
 \det\left[1-
T_{N_2}\left[f^{(2)}\right]^{-1} H\left[f^{(2)}\right]T_{N_1}\left[\widetilde{f}^{(1)}\right]^{-1} H\left[\widetilde{f}^{(1)}\right]P_{N_2}
\right]\,.
\label{det_A}
\end{equation}
We have used here that $Q_N T_N[f]Q_N=T_N[\widetilde{f}]$. It is important that
the last determinant in Eq.~\ref{det_A} 
has a finite limit as $N_0$ and $N_1$ go to infinity, 
\begin{equation}
 C=\det\left[1-
T\left[f^{(2)}\right]^{-1} H\left[f^{(2)}\right]T\left[\widetilde{f}^{(1)}\right]^{-1} H\left[\widetilde{f}^{(1)}\right]
\right]\,. 
\end{equation}
Here $T[f]$ stands for semi-infinite Toeplitz matrix with symbol $f$. We can now
use simple properties of semi-infinite 
Toeplitz and Hankel matrices \cite{Widom76}
\begin{eqnarray}
T[\phi]T[\psi]+H[\phi]H[\widetilde{\psi}]=T[\phi\psi]\\
 H[\phi]T[\widetilde{\psi}]+T[\phi]H[\psi]=H[\phi\psi]
\end{eqnarray}
to bring the expression for constant $C$ to the form
\begin{equation}
 C=\det\left[T\left[f^{(2)}\right]^{-1}T\left[f^{(2)}/f^{(1)}\right]
T\left[1/f^{(1)}\right]^{-1}\right]\,.
\label{CFinal}
\end{equation}
On the other hand, one can recast the Szeg\H{o} result (\ref{SzegoClassical})
into the form\cite{Widom76}
\begin{equation}
 \det T_N[f]=e^{NV_0}\det\left[T[f^{-1}]T[f]\right]\,.
\label{SzegoWidom}
\end{equation}
Combining now Eqs. (\ref{CFinal}) and (\ref{SzegoWidom}) we get for the determinant of interest
\begin{equation}
\det A_{N_1, N_2}=e^{N_1 V^{(1)}_{0} +N_2 V^{(2)}_{0}}\\\times
\det\left[T\left[1/f^{(2)}\right]T\left[f^{(2)}/f^{(1)}\right]
T\left[f^{(1)}\right]\right]\,,
\end{equation}
where $V_0^{(i)}$ is the zero Fourier component of $f^{(i)}$. 
The derivation above can be immediately generalized to the case of $\delta(t)$ having arbitrary number of steps,  Eq.(\ref{multipulse-delta_app}). 
The resulting determinant reads
\begin{equation}
 \det A_{N_1, \ldots N_{k}}\left[f^{(1)},\ldots f^{(k)}\right]=e^{\sum_{i=1}^kN_iV^{(i)}_{0}}\\\times
\det\left[T\left[g^{(k)}\right]\ldots T\left[g^{(0)}\right]\right]
\label{DetWidomForm}
\end{equation}
with
\begin{eqnarray}
 g^{(i)}=\frac{f^{(i+1)}}{f^{(i)}}\,, \qquad i=0, \ldots k\,\,\, {\rm and} \\ 
f^{(0)}=f^{(k+1)}\equiv 1\,.
\end{eqnarray}
Note that by definition
\begin{equation}
\prod_{i=1}^k g^{(i)}=1\,.
\label{gProd} 
\end{equation}

Equation (\ref{DetWidomForm}) allows to reduce the problem of asymptotic
behavior of $\det A_{N_1, \ldots N_k}$ to the evaluation  
of a determinant of a product of semi-infinite Toeplitz matrices. This is a
crucial step forward because it enables us to apply 
the Winer-Hopf method to the problem. Specifically, let us introduce  the
Wiener-Hopf decomposition of the functions $g^{(i)}$,
\begin{equation}
 g^{(i)}(z)=e^{U^{(i)}_+(z)}e^{U^{(i)}_0} e^{U^{(i)}_-(z)}\,.
\end{equation}
Here $U^{(i)}_{+(-)}(z)$ are analytic inside (outside) the unit circle.  The analytic properties of $U_{\pm}^{(i)}$ 
allow us to write
\begin{equation}
 T[g^{(i)}]=e^{U^{(i)}_0}e^{T\left[U^{(i)}_-\right]}e^{T\left[U^{(i)}_+\right]}\,.
\end{equation}
Thus 
\begin{equation}
 \det A_{N_1, \ldots N_{k}}\left[f^{(1)},\ldots f^{(k)}\right]=e^{\sum_{i=1}^kN_iV^{(i)}_{0}}\\\times
\det\left[e^{U^{(k)}_0}e^{T\left[U^{(k)}_-\right]}e^{T\left[U^{(k)}_+\right]}\ldots e^{U^{(0)}_0}e^{T\left[U^{(0)}_-\right]}e^{T\left[U^{(0)}_+\right]}\right]\,.
\label{DetExponentForm}
\end{equation}
Note that due to the relation (\ref{gProd}) 
\begin{eqnarray}
 \sum_{i=1}^k U^{(i)}_0=\sum_{i=1}^kT\left[U^{(i)}_\pm\right]=0 \,,
\end{eqnarray}
and one could naively conclude that the determinant in (\ref{DetExponentForm})
is equal to unity. This is not true, however, due to the infinite size of the
matrices involved. Instead, one should make use of the
formula \cite{Ehrhardt2003}
\begin{equation}
 \det e^{U_R}\ldots e^{U_1}=\exp\left[\frac12\sum_{1\leq j<i\leq R} \tr\left[U_i, U_j\right]\right]
\label{DetGeneral}
\end{equation}
valid for arbitrary set of operators $U_i$ satisfying $\sum_{i=1}^R U_i=0$. Here
$[*,*]$ stands for the commutator. Applying (\ref{DetGeneral}) to
(\ref{DetExponentForm}), we obtain
\begin{equation}
 \det A_{N_1, \ldots N_{k}}\left[f^{(1)},\ldots f^{(k)}\right]=e^{\sum_{i=1}^kN_iV^{(i)}_{0}}\\\times
\exp\left[\sum_{0\leq i<j\leq k}\tr\left[T\left[U^{(j)}_+\right], T\left[U^{(i)}_-\right]\right]\right]\,.
\end{equation}
Finally, evaluating the traces one finds
\begin{equation}
 \det A_{N_1, \ldots N_{k}}\left[f^{(1)},\ldots f^{(k)}\right]=e^{-\sum_{i=0}^kU^{(i)}_{0}\sum_{j=1}^i N_j}\\\times
  \exp \left[-\sum_{0\leq i<j\leq k}\sum_{q=1}^{\infty}
qU_{q}^{(j)}U_{-q}^{(i)}\right]\,.
\label{SzegoGeneral}
\end{equation}
Equation (\ref{SzegoGeneral}) is the generalization of the Szeg\H{o} limit
theorem to the case of determinants with a multi-step $A(t)$  [and smooth
$B(\epsilon)$] and constitutes the main result of this subsection. In the next
subsection we will analyze its implications for the case when $B(\epsilon)$ is
not smooth but rather has multiple steps.

\subsection{Multiple singularities in $A(t)$ and in
$B(\epsilon)$:\ \ Further generalization of Fisher-Hartwig conjecture}
\label{subsec:gen-Fisher-Hartwig}

We consider now the case when not only the phase $\delta(t)$ but also the
distribution function $n(\epsilon)$ has multiple step-like singularities, 
Eq.~(\ref{multistep-distribution_app}).
We will apply the generalized Szeg\H{o} formula Eq.~(\ref{SzegoGeneral})
derived above to this situation and then cut off the resulting logarithmic
divergencies. While this is not a mathematically rigorous procedure, we should
get in this way correct results for power-law exponents (apart from the
summation over branches of the logarithms), cf.
Sec.~\ref{S:SzegoVSFisherHartwig}. 
\begin{figure}
 \includegraphics[width=230pt]{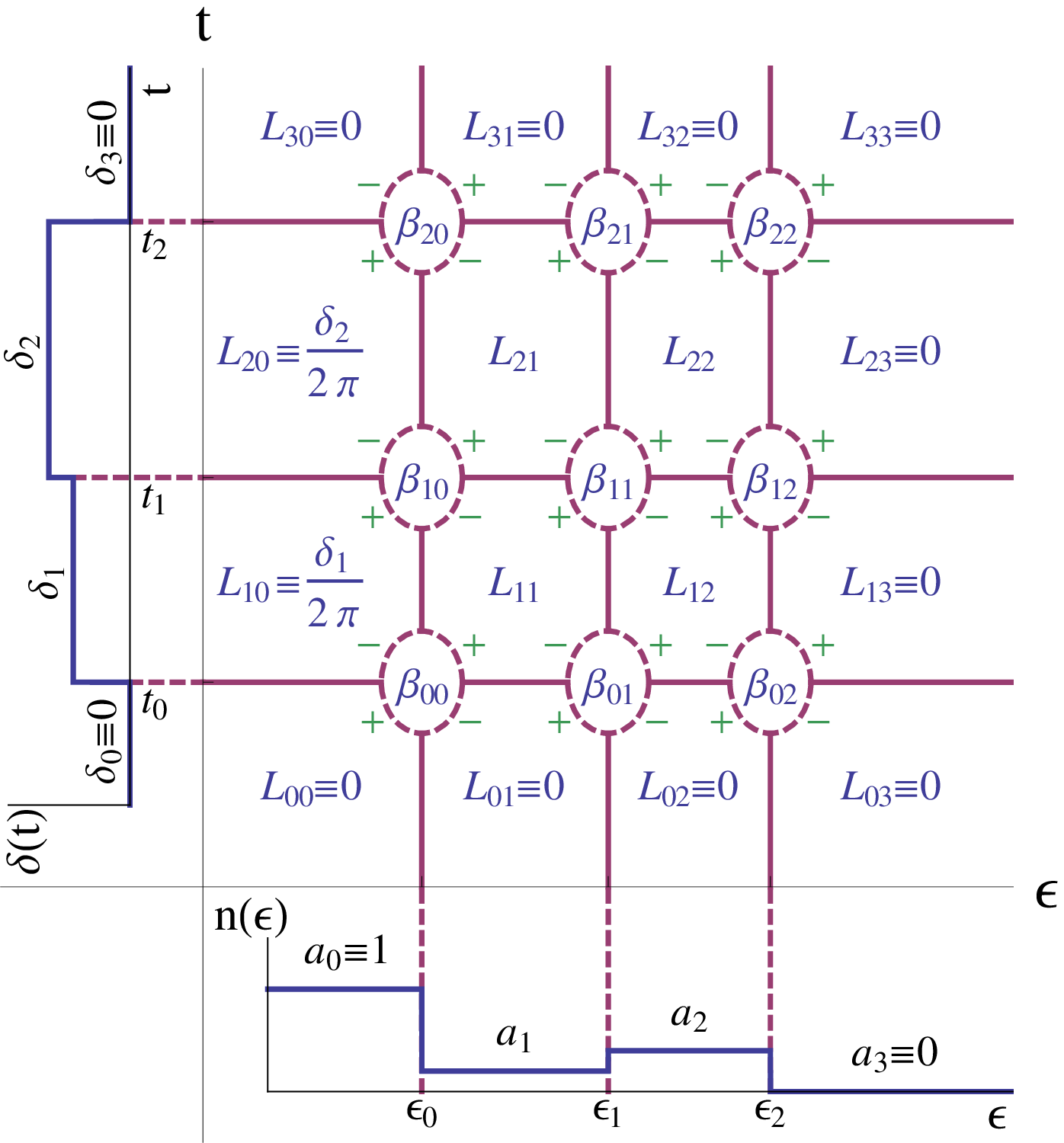}
\caption{\small Construction of the numbers $\beta_{lp}$ determining the asymptotic behavior of the Fredholm determinant. 
Each of the coefficients $\beta_{lp}$ can be represented as $\beta_{ij}=L_{i, j}+L_{i+1,j+1}-L_{i+1, j}-L_{i, j+1}$ with
$L_{ij}=\frac{i}{2\pi }\log\left[1+(e^{-i\delta_i}-1)a_j\right]$.
}
\label{BettaGraph}
\end{figure} 

The relevant functions $g^{(i)}$ are now given by
\begin{equation}
 g^{(i)}(z)=e^{U^{(i)}_0}z^{\sum_{j=0}^m\beta_{i, j}} \prod_{j=0}^{m}g_{z_j, \beta_{ij}}(z)z_j^{-\beta_{ij}}
\end{equation}
with
\begin{equation}
 U^{(i)}_0= \sum_{j=0}^m\beta_{i, j}\left(\ln z_j-i\pi\right)
\end{equation}
and
\begin{equation}
 \beta_{ij}=\frac{1}{2\pi i}\left(
\ln \frac{1+(e^{-i\delta_{i+1}}-1)a_j}{1+(e^{-i\delta_{i+1}}-1)a_{j+1}}\right.\\\left.
-\ln \frac{1+(e^{-i\delta_i}-1)a_j}{1+(e^{-i\delta_i}-1)a_{j+1}}
\right)\\\equiv L_{i, j}+L_{i+1,j+1}-L_{i+1, j}-L_{i, j+1}\,.
\end{equation}
Here  we have intriduced the notation
\begin{equation}
 L_{ij}=\frac{i}{2\pi }\log\left[1+(e^{-i\delta_i}-1)a_j\right]\,.
\label{LDef}
\end{equation}
The coefficients $\beta_{i,j}$ can be assigned to the singular points
$(\epsilon_j, t_i)$ in $(\epsilon, t)$ plane, as  illustrated in Fig.
\ref{BettaGraph}.  

Just as in Sec. \ref{S:SzegoVSFisherHartwig} coefficients $\beta_{ij}$ control the asymptotic behavior of $U^{(i)}_q$
at $q\rightarrow\infty$,   
\begin{equation}
 U^{(i)}_q\sim \frac{1}{q}\sum_{j=0}^m\beta_{ij}\,.
\end{equation}
Applying now Eq.~(\ref{SzegoGeneral})
with  the cutoff procedure discussed in Sec.\ref{S:SzegoVSFisherHartwig}
and representing the result in terms of physical variables,  
we find an asymptotic behavior of the determinant of interest
\begin{equation}
 \Delta\left[\delta(t), n(\epsilon)\right]=C[\beta]\exp\left[-i\sum_{i,
j}t_i\beta_{ij}(\epsilon_j-\Lambda)\right]\prod_{0\leq i'<i\leq
k}\left[\Lambda(t_i-t_{i'})\right]^{\sum_{j=0}^m
\beta_{ij}\beta_{i'j}}\prod_{j'<j}\left(\frac{\Lambda}{\epsilon_j-\epsilon_{j'}}\right)^{-\sum_{i=0}^k
\beta_{ij'}\beta_{ij}}\,,
\label{DeltaGeneral}
\end{equation}
where  we have  used  the condition 
\begin{equation}
 \sum_{i=0}^k\beta_{ij}=0\,.
\end{equation}

In full analogy with the case of Toeplitz determinants with Fisher-Hartwig
singularities, see Sec.~\ref{S:SzegoVSFisherHartwig},
Equation (\ref{DeltaGeneral}) suffers from the ambiguity in the choice of
$\beta_{ij}$. Pursuing this analogy, we conclude that the correct 
generalization of (\ref{DeltaGeneral}) 
involves a  summation over all possible choices of $\beta_{ij}$,
\begin{equation}
 \Delta\left[\delta[t], n(\epsilon)\right]=\sum_{\mathrm br} C[\beta]\exp\left[-i\sum_{i, j}t_i\beta_{ij}(\epsilon_j-\Lambda)\right]\prod_{0\leq i'<i\leq k}\left[\Lambda(t_i-t_{i'})\right]^{\sum_{j=0}^m
\beta_{ij}\beta_{i'j}}\prod_{j'<j}\left(\frac{\Lambda}{\epsilon_j-\epsilon_{j'}}\right)^{-\sum_{i=0}^k
\beta_{ij'}\beta_{ij}}\,.
\label{DeltaGeneralSum}
\end{equation}
The summation over $\beta_{ij}$ here can be understood as
summation over  all possible branches of logarithms in the definition of
$L_{ij}$, Eq. (\ref{LDef}), 
with the additional constraints imposed [cf. Eq.(\ref{nConstraint})]
\begin{eqnarray}
 L_{0j}=L_{k+1, j}=L_{i, m+1}\equiv 0 \ \  {\rm and } \ \ L_{i,
0}\equiv\frac{\delta_i}{2\pi}\,.
\end{eqnarray}

A particular case of (\ref{DeltaGeneralSum}) is the zero temperature
determinant
\begin{equation}
 \Delta\left[\delta(t), T=0\right]=C_0[\beta]\exp\left[i\Lambda\sum_{i}t_i\beta^{(0)}_{i0}\right]\\\prod_{0\leq i'<i\leq k}\left[\Lambda(t_i-t_{i'})\right]^{\beta^{(0)}_{i0}\beta^{(0)}_{i'0}}
\label{Delta0}
\end{equation}
with $ \beta^{(0)}_{i0}=(\delta_i-\delta_{i+1})/2\pi$. This result can also be
checked by a direct calculation using the fact that at equilibrium the
determinant is a gaussian functional of $\delta(t)$.

Equation (\ref{DeltaGeneralSum}) can be represented in a more symmetric way. 
Employng the condition  
\begin{equation}
 \sum_{j=0}^m\beta_{ij}=\frac{\delta_i-\delta_{i+1}}{2\pi}=\beta_{i0}^{(0)}
\end{equation}
(valid for arbitrary multi-step distribution) and  
combining it with  (\ref{DeltaGeneralSum}) and 
(\ref{Delta0}) 
we arrive at the (cutoff independent) asymptotic of the normalized determinant 
$\overline{\Delta}[\delta(t), n(\epsilon)] = \Delta[\delta(t),n(\epsilon)] /
\Delta[\delta(t), T=0]$, 
\begin{equation}
 \overline{\Delta}[\delta(t), n(\epsilon)]= \sum_{\mathrm{br}}C[\beta] \exp\left[-i\sum_{i\,,j}t_i\beta_{ij}\epsilon_j\right]
\prod_{0\leq i'<i\leq k}\;\prod_{0\leq j'<j\leq m}\left[(t_i-t_{i'})(\epsilon_j-\epsilon_{j'})\right]^{\gamma_{ii';jj'}}\,,
\label{MainResFinal}
\end{equation}
where we have  introduced the exponents 
\begin{equation}
 \gamma_{ii', jj'}=-\beta_{ij}\beta_{i'j'}-\beta_{ij'}\beta_{i'j}\,.
\end{equation}

Let us now  discuss the applicability of the long time  asymptotic expansion.
Throughout the derivations of Eq.~(\ref{MainResFinal}),  it was assumed that the
area of all the rectangles in the time-energy plane is large, i.e.,  for any
$t_i$, $t_{i'}$, $\epsilon_j$ and $\epsilon_{j'}$ 
\begin{equation}
 \left(\epsilon_i-\epsilon_{i'}\right)\left(t_j-t_{j'}\right)\gg 1 \,, \qquad i\neq i'\,, \qquad j\neq j'\,.
\label{AsymptoticCondition}
\end{equation}
A straightforward extension of the cutoff procedure  shows that
asymptotics (\ref{MainResFinal}) remains valid  even  if for some sets
$(i,i',j,j')$ the condition (\ref{AsymptoticCondition}) is not satisfied,
provided one drops the corresponding factors from   the product
(\ref{MainResFinal}).

\section{Factorization of the determinant for $\delta(t)$ consisting of several
remote pulses}

%%%%%%%%%%%%%%%%%%%%%%%%%%%%%%%%%%%%%%%%%%%%%%%%%%%%%%%%%%%%%%%%%%%%
\begin{figure}
 \includegraphics[width=230pt]{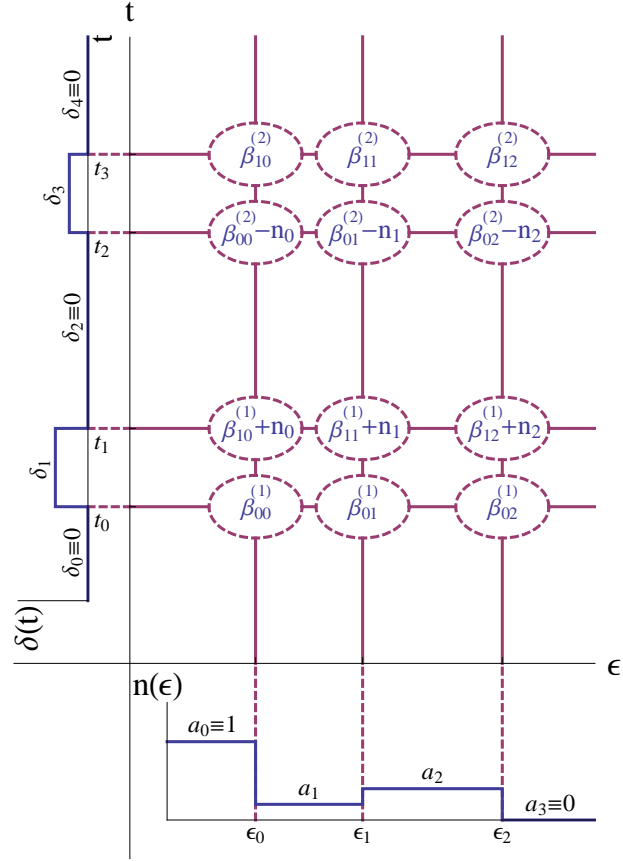}
\caption{\small The set of number $\beta_{ij}$ corresponding to the counting field $\delta(t)$, 
Eq.~(\ref{fourstep-delta}),  in terms of $\beta_{ij}^{(1)}$ and
$\beta_{ij}^{(2)}$ corresponding to individual pulses  comprising $\delta(t)$.
}
\label{BettaGraphDecoupling}
\end{figure} 
%%%%%%%%%%%%%%%%%%%%%%%%%%%%%%%%%%%%%%%%%%%%%%%%%%%%%%%%%%%%%%%%%%%%%

In this section we discuss the determinant $\Delta[\delta(t), n(\epsilon)]$ for
the phase  $\delta(t)$  consisting of several rectangular pulses separated by
large time intervals. We show that in this limit the determinant is given by a
product of Toeplitz determinants
corresponding to individual   rectangular  pulse. 
This question is relevant for the calculation  of correlation
functions in Luttinger liquid within generic  boundaries (and in particular for
``sharp boundary'' model).  In our current analysis we rely  on
Eq.(\ref{DeltaGeneralSum}).

Let us consider counting field $\delta(t)$ given by (see Fig.~\ref{BettaGraphDecoupling})
\begin{equation}
 \label{fourstep-delta}
\delta(t) = \left\{
\begin{array}{ll}
0 \equiv \delta_0 \,, & \qquad t < t_0 \\
\delta_1\,, & \qquad t_0 < t <t_1 \\
 0\,, & \qquad t_{1} < t <t_2 \\
 \delta_3\,, & \qquad t_2 < t <t_3 \\
0 \equiv \delta_{4}\,, & \qquad t_3 < t\,.  
\end{array}
\right.
\end{equation}
Our aim is to show that as $t_2-t_1$ goes to infinity, $\Delta[\delta(t),
n(\epsilon)$ decouples into the product 
$\Delta[\delta^{(1)}(t), n(\epsilon)]\Delta[\delta^{(2)}(t), n(\epsilon)]$, with
$\delta^{(1)}(t)$ and $\delta^{(2)}(t)$ 
being the two rectangular pulses constituting $\delta(t)$. 

Let us introduce coefficients $\beta^{(l)}_{ij}$ with $l=1, 2$ corresponding to
the phases $\delta^{(1)}(t)$ and  
$\delta^{(2)}(t)$.  Figure \ref{BettaGraphDecoupling} demonstrates the
connection between the coefficients  
$\beta_{ij}$ corresponding to $\delta(t)$ and $\beta^{(l)}_{ij}$. Here we have
introduced a set of integers $n_j$  
satisfying $\sum_j n_j=0$.  Let us consider a particular term in the sum
(\ref{DeltaGeneralSum}) corresponding to the given set of coefficients
$\beta^{(l)}_{ij}$ and set of integers $n_j$. The dependence of this term on
$t_2-t_1$ at $t_2-t_1\rightarrow \infty$ is easily found to be   
\begin{equation}
 (t_2-t_1)^{\sum_{j=0}^m\sum_{i, i'=1, 2}\beta_{ij}^{(1)}\beta_{i'j}^{(2)}}=
(t_2-t_1)^{-\sum_{j=0}^mn_j^2}\,\,\,.
\end{equation}
We see that only terms of the sum (\ref{DeltaGeneralSum}) characterized by
$n_j=0$, for $j=0, \ldots m$   survive  the limit  $t_2-t_1\rightarrow\infty$.
Thus we find that in the long-time limit the determinant in question factorizes
into a product of Toeplitz determinants
\begin{equation}
 \Delta[\delta(t), n(\epsilon)]=\Delta[\delta^{(1)}(t),
n(\epsilon)]\Delta[\delta^{(2)}(t), n(\epsilon)]\,, \;  
t_2-t_1\rightarrow \infty\,.
\end{equation}
This factorization  was proposed in Ref.\cite{GGM_short2010,GGM_long2010}  on the  basis of 
physical arguments and employed to calculate single-particle Green
functions.
In the main text of the present paper we also use an analogous factorization
for the case of two-particle correlation functions.

\section{Numerical check for the asymptotics (\ref{MainResFinal}) of
determinants with multiple $t$ and $\epsilon$ discontinuities}

%%%%%%%%%%%%%%%%%%%%%%%%%%%%%%%%%%%%%%%%%%%%%%%%%%%%%%%%%%%%%%%%%%%%%%%
\begin{figure}
\includegraphics[width=410pt]{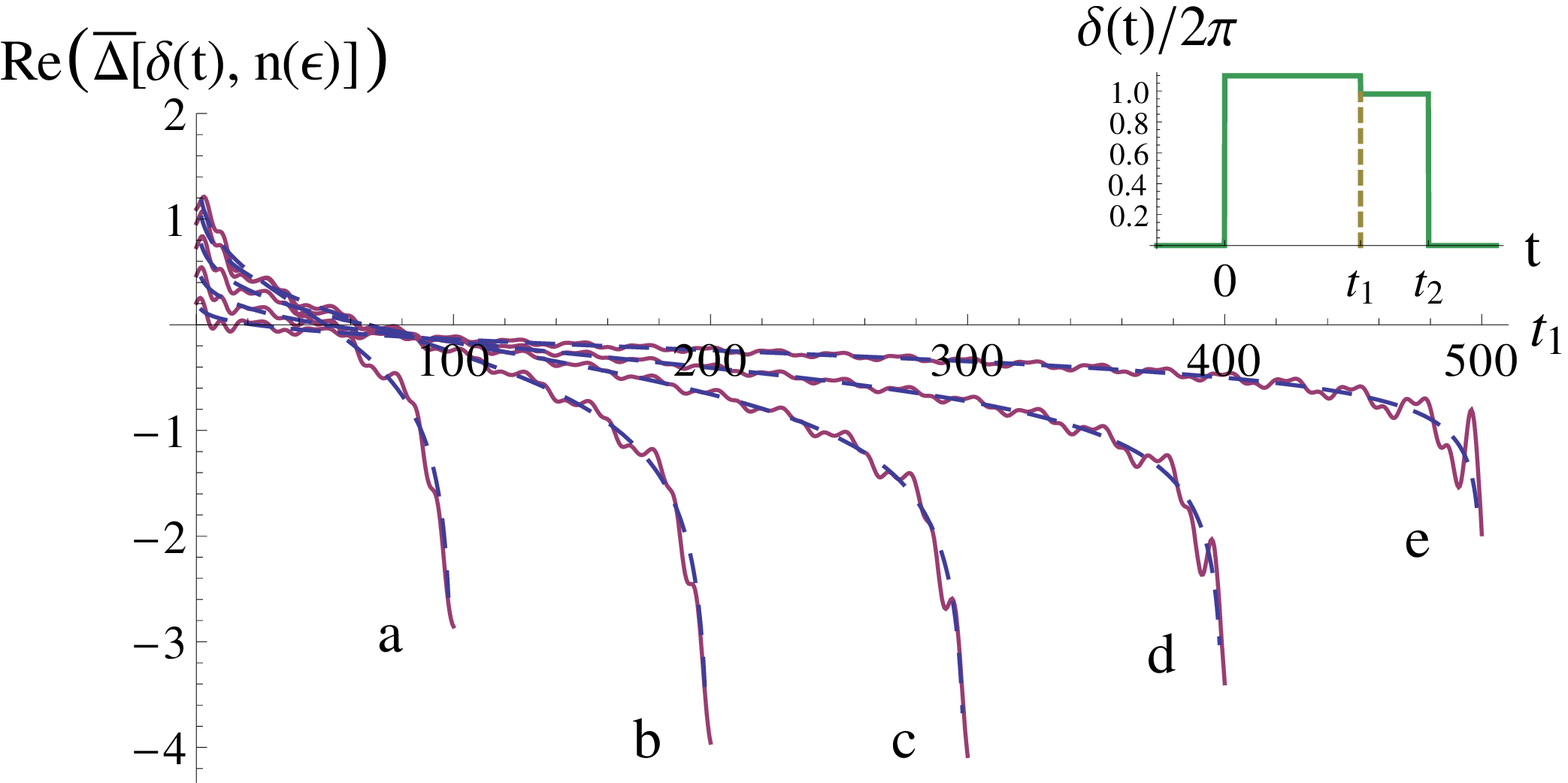}
 \includegraphics[width=410pt]{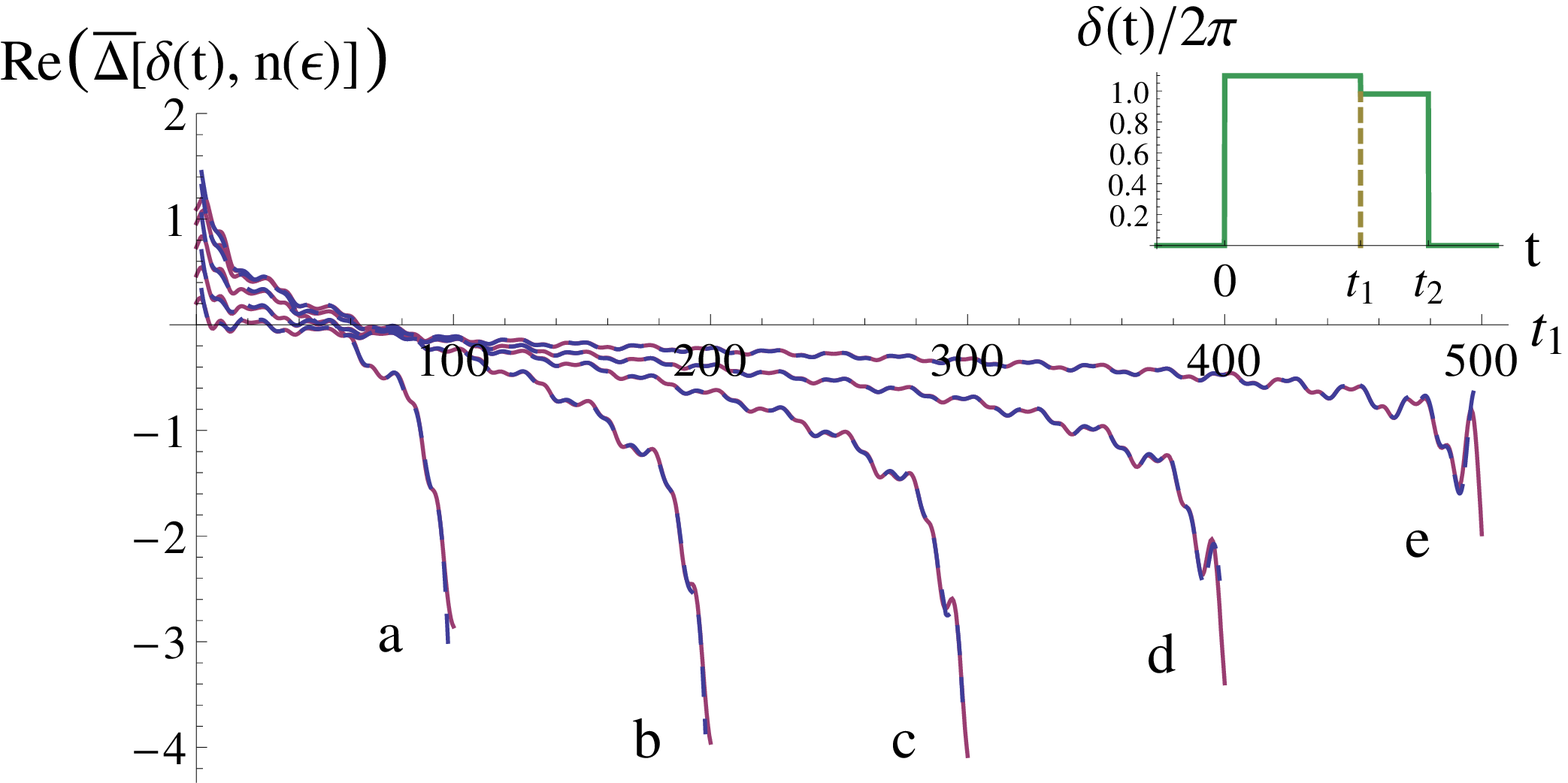}
\caption{\small The real part of the determinant $\overline{\Delta}[\delta(t),
n(\epsilon)]$ for triple-step counting field $\delta(t)$ (see inset) and
triple-step distribution $n(\epsilon)$. The solid blue line on both graphs
represents the results of numerical evaluation of Fredholm determinant. Each
curve corresponds to the fixed time $t_2$ ($t_2=100,  200,\ldots 500$ for curves
mark by letters from ``a'' to ``e'') while $t_1$ changes from $0$ to $t_2$.
{\it Upper panel.} Dashed lines show the fit of numerical result by expression
(\ref{Delta1}) with coefficients $C_1$,  
$C_2$  and $C_3$ considered as fitting parameters. The same set of $C_i$ is used
for all curves.  
{\it Lower panel.} Dashed lines show the fit of numerical result with the
correction terms (\ref{Delta2}) taken into  account. The oscillations of the
determinant with time $t_1$ are now correctly reproduced.
}
\label{AsymptoticCheck}
\end{figure}
%%%%%%%%%%%%%%%%%%%%%%%%%%%%%%%%%%%%%%%%%%%%%%%%%%%%%%%%%%%%%%%%%%%%%%%%%%%%

In this section we present a numerical verification of the
asymptotic formula (\ref{MainResFinal}) for determinants with multiple
discontinuities both in $A(t)$ and in $B(\epsilon)$. We calculate numerically
the normalized Fredholm determinant $\overline{\Delta}[\delta(t), n(\epsilon)]$
with a triple-step counting field (see inset of Fig.\ref{AsymptoticCheck})
\begin{equation}
 \delta(t) = \left\{
\begin{array}{ll}
0  \,, & \qquad t < 0 \\
\delta_1=1.1\times 2\pi\,, & \qquad 0 < t <t_1 \\
 \delta_2=0.98\times 2\pi\,, & \qquad t_1 < t <t_2 \\
0 \,, & \qquad t_2 < t\,, 
\end{array}
\right.
\label{delta_triple_step}
\end{equation}
as a function of $t_1$ and $t_2$ and compare results to the predictions of
Eq.(\ref{MainResFinal}). The energy distribution is taken to be 
\begin{equation}
 n(\epsilon) = \left\{
\begin{array}{ll}
1  \,, & \qquad \epsilon < \epsilon_0 \\
0\,, & \qquad \epsilon_0 < \epsilon <\epsilon_1 \\
1\,, & \qquad \epsilon_1<\epsilon<\epsilon_2\\
0 \,, & \qquad \epsilon_2 < \epsilon\,,  
\end{array}
\right.
\end{equation}
 with $\epsilon_0=-3/4$, $\epsilon_1=-1/2$ and $\epsilon_2=1/4$.  
The numerical procedure used here is based on the time discretization and is
analogous to the one employed in Ref.~\cite{Protopopov2012} in course of
studies of Toeplitz determinants.

Taking into account that $\delta_1$ and $\delta_2$ in (\ref{delta_triple_step})
are close to $2\pi$, one infers the following  three  dominant contributions to
the sum~(\ref{MainResFinal}): 
\begin{equation}
 \overline{\Delta}^{(1)}=\left(C_1e^{-i\epsilon_0
t_2}+C_2e^{-i\epsilon_2t_2}\right)t_1^{\gamma^{(1)}_1}t_2^{\gamma^{(1)}_2}
(t_2-t_1)^{\gamma^{(1)}_{12}}+C_3 e^{-i\epsilon_1 t_2} 
t_1^{\gamma^{(2)}_1}t_2^{\gamma^{(2)}_2}(t_2-t_1)^{\gamma^{(2)}_{12}}
\label{Delta1}
\end{equation}
Here we have absorbed the dependence of the determinant on energies $\epsilon_j$
into the coefficients $C_i$. 

The exponents in (\ref{Delta1}) are given by 
\begin{eqnarray}
 \gamma^{(1)}_1=-\frac{1}{2\pi^2} (\delta_1-2\pi) (\delta_1-\delta_2)\,,
\label{gamma1}
\\
 \gamma^{(1)}_2=\frac{1}{2\pi^2} (-\delta_1\delta_2+2\pi\delta_1+2\pi\delta_2-4\pi^2)\,,
\label{gamma2}
\\
 \gamma^{(1)}_{12}=\frac{1}{2\pi^2} (\delta_2-2\pi) (\delta_1-\delta_2)\,,
\label{gamma3}
\\
 \gamma^{(2)}_1=-\frac{1}{2\pi^2} (\delta_1-4\pi) (\delta_1-\delta_2)\,,
\label{gamma4}
\\
 \gamma^{(2)}_2=\frac{1}{2\pi^2} (-\delta_1\delta_2+4\pi\delta_1+4\pi\delta_2-12\pi^2)\,,
\label{gamma5}
\\
 \gamma^{(2)}_{12}=\frac{1}{2\pi^2} (\delta_2-4\pi) (\delta_1-\delta_2)\,.
\label{gamma6}
\end{eqnarray}
A characteristic feature of the exponents (\ref{gamma1})---(\ref{gamma6}) is
their smallness at  
$\delta_1\approx\delta_2\approx2\pi$. 

The upper panel of Fig.\ref{AsymptoticCheck} shows the real part of the
determinant $\overline{\Delta}[\delta(t), n(\epsilon)]$.  The solid blue lines
on both graphs represent the results of numerical evaluation of Fredholm
determinant. Each curve corresponds to the fixed time $t_2$ ($t_2=100,
200,\ldots 500$ for curves mark by letters from ``a'' to ``e'', respectively)
while $t_1$ changes from $0$ to $t_2$. Dashed lines show the fit of numerical
result by expression (\ref{Delta1}) with coefficients $C_1$, $C_2$  and $C_3$
considered as fitting parameters. The same set of $C_i$ is used for all curves. 
We see that the fit reproduces correctly the overall behavior of the determinant
but the oscillations with time $t_1$ are not captured in this approximation.
An analysis of (\ref{MainResFinal}) shows that for chosen parameters  the
dominant correction to (\ref{Delta1}) is given by 
\begin{equation}
 \overline{\Delta}^{(2)}=\left(C_4e^{-i\epsilon_0 t_2-i\epsilon_2 t_1}+C_5e^{-i\epsilon_2 t_1-i\epsilon_1t_2}\right)t_1^{\gamma^{(3)}_1}t_2^{\gamma^{(3)}_2}(t_2-t_1)^{\gamma^{(3)}_{12}}
\label{Delta2}
\end{equation}
with
\begin{eqnarray}
 \gamma^{(3)}_1=\frac{1}{2\pi^2}\delta_1(\delta_2-\delta_1)+\frac{1}{\pi}(3\delta_1-2\delta_2)-3\,,\\
 \gamma^{(3)}_2=-\frac{1}{2\pi^2}\delta_1\delta_2+\frac{1}{\pi}(2\delta_2+\delta_1)-3\,,\\
 \gamma^{(3)}_{12}=\frac{1}{2\pi^2}\delta_2(\delta_1-\delta_2)-\frac{1}{\pi}
\delta_1+1\,.
\end{eqnarray}
Taking the correction (\ref{Delta2}) into account, we obtain a fit to
numerical data (bottom panel of  Fig.~\ref{AsymptoticCheck})
which correctly captures  the oscillation in time.  The resulting agreement is
essentially perfect. 
We thus conclude that our conjecture, Eq.  (\ref{MainResFinal}), is fully
supported by the numerical simulations.  
\end{widetext}


\begin{thebibliography}{99}
\bibitem{CommentLocalLog} If the functions $A(t)$ and/or $B(\epsilon)$ are not
constant between the singularities, one has to take the values of $L$'s near
the corresponding corner.
\end{thebibliography}
\end{document}